\documentclass[11pt, a4paper, copyright, gdm]{google}

\usepackage[authoryear, sort&compress, round]{natbib}
\usepackage{multirow}
\usepackage{array}
\usepackage{placeins}   %
\usepackage{listings}
\usepackage{upquote}
\ifx\XeTeXrevision\undefined\else
\fi
\usepackage[capitalize,noabbrev]{cleveref}
\crefname{lstlisting}{Listing}{Listings}
\Crefname{lstlisting}{Listing}{Listings}
\definecolor{codegreen}{rgb}{0,0.55,0}
\definecolor{codegray}{rgb}{0.5,0.5,0.5}
\definecolor{codepurple}{rgb}{0.52,0,0.72}
\definecolor{codeblue}{rgb}{0.0,0.1,0.65}
\newcommand{\ours}{Qwen3.5-9B\,+\,RL}   %
\newcommand{\sftm}{Qwen3.5-9B\,+\,SFT}   %

\definecolor{tabRL}{HTML}{E8F1FC}    %
\definecolor{tabPanel}{HTML}{EEF2F7} %
\definecolor{gainc}{HTML}{1B7F3B}    %
\definecolor{lossc}{HTML}{B3261E}

\newcommand{\ci}[1]{{\fontsize{5.5}{6}\selectfont\color{black!50}\,$\pm$#1}}   %
\usepackage[most]{tcolorbox}
\definecolor{promptSys}{HTML}{1A4D8F}
\definecolor{promptFb}{HTML}{0B6E69}
\definecolor{promptJudge}{HTML}{5B3E96}
\definecolor{promptPh}{HTML}{B3541E}
\newtcolorbox{promptbox}[2][promptSys]{enhanced, breakable, colback=#1!4!white, colframe=#1, coltitle=white,
  fonttitle=\sffamily\bfseries\small, title={#2}, boxrule=0.6pt, arc=2.5pt, left=7pt, right=7pt, top=5pt,
  bottom=5pt, toptitle=2pt, bottomtitle=2pt, before skip=6pt, after skip=8pt,
  before upper={\scriptsize\ttfamily\raggedright\setlength{\parskip}{0pt}\colorlet{promptHead}{#1}}}
\newcommand{\promptline}[1]{\par\noindent #1\par}
\newcommand{\promptitem}[1]{\par\noindent\hangindent=1.6em\hangafter=1 #1\par}
\newcommand{\promptgap}{\par\vspace{0.3em}}
\newcommand{\prompthead}[1]{\par\noindent{\sffamily\bfseries\color{promptHead}#1}\par}
\newcommand{\ph}[1]{{\color{promptPh}\itshape\{#1\}}}


\uselogo{}

\title{Learning In-Context Adaptivity for LLMs through Continual Gameplay}

\correspondingauthor{jjl7199@psu.edu, lizun@google.com}

\reportnumber{0001} %

\author[1]{Jiale Liu}
\author[2]{Zun Li}
\author[2]{John Schultz}

\affil[1]{Penn State University, Work done during an internship at Google DeepMind}
\affil[2]{\thepa{}{}}

\begin{abstract}
LLM agents are increasingly deployed in long-horizon tasks where they must explore, gather experience, and adapt their decisions accordingly.
We name this capability \emph{in-context adaptivity}: improving decisions across episodes from experience accumulated in context.
We propose to evaluate it in repeated imperfect-information games against unknown opponents.
In these games, fixed game-theoretically optimal policies cannot achieve maximum reward, whereas adapting to the opponent increases returns across games.
We design a paired protocol that evaluates each agent with and without memory under identical conditions to isolate \emph{experience gain} from single-game skill.
Evaluation shows that memory brings little performance gain for open-weight models and inconsistent gains for frontier models.
Our analysis of the agents' summaries and actions identifies a \emph{memory-doing gap} in which models accurately record opponent tendencies yet rarely depart from default play.
To close this gap, we construct a pool of exploitable opponents and post-train Qwen3.5-9B with supervised fine-tuning followed by trajectory-level reinforcement learning.
We optimize full-match return across context segments so summary generation and action selection are trained together.
On unseen opponents, our 9B RL policy doubles experience gain over supervised fine-tuning in Leduc Hold'em and increases it fivefold in No-Limit Hold'em.
Across 35 opponents, our policy matches or surpasses frontier models, generalizes beyond its training horizon, and retains most of its gain on summaries from frontier models.
\end{abstract}

\begin{document}

\maketitle
\suppressfloats[t]   %

\section{Introduction}
\label{sec:intro}

\begin{figure*}[tb]
\centering
\includegraphics[width=\textwidth]{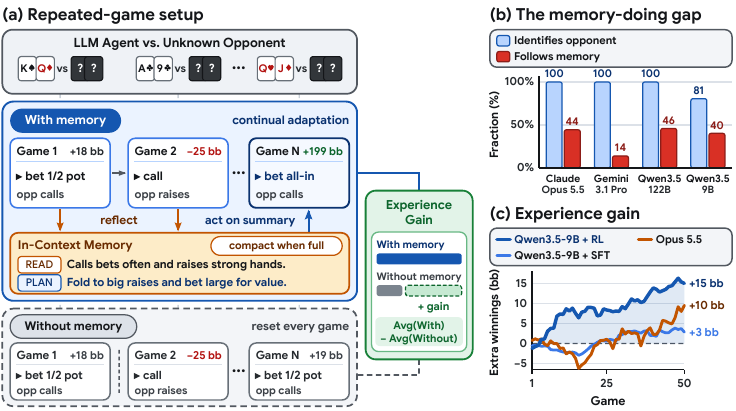}
\vspace{-0.4em}
\caption{\textbf{Measuring and training in-context adaptivity in repeated imperfect-information games.}
\textbf{(a)}~Comparing continual play with memory against play without memory under identical conditions isolates experience gain $\Delta_{\text{exp}}$.
\textbf{(b)}~Share of final summaries identifying an exploitable opponent tendency versus decisions where memory switches the action to the best response in Leduc Hold'em.
\textbf{(c)}~Cumulative extra winnings from memory over play without memory across unseen No-Limit Hold'em opponents.}
\vspace{-1.0em}
\label{fig:teaser}
\end{figure*}

As language models grow more capable, they are increasingly deployed as autonomous agents that operate over long horizons \citep{wang2024survey,anthropic2025claudecode,openai2025codex}.
This progress holds the promise of continual learning agents that keep improving from their own experience after deployment \citep{gao2025survey,fang2025comprehensive,singh2026plasticity}.
To do so in an unfamiliar environment, an agent must explore, gather observations, and adapt its decisions as experience accumulates.
We refer to this capability as \emph{in-context adaptivity}, the ability to improve decisions across sequential episodes from experience accumulated in context.
Modern agent systems routinely maintain past interactions through reflection and periodic summarization \citep{shinn2023reflexion,packer2023memgpt,zhang2025memorysurvey,hu2025memorysurvey}, yet whether language models actually turn accumulated context into better decisions remains unclear \citep{he2026memoryarena,zhao2026amabench}.

To evaluate in-context adaptivity, we study repeated imperfect-information games against unknown opponents, focusing on Leduc Hold'em \citep{southey2005bayes} and Heads-Up No-Limit Texas Hold'em \citep{moravcik2017deepstack,brown2018superhuman}.
In this setting, fixed game-theoretically optimal policies cannot achieve maximum reward, whereas inferring and exploiting an opponent's tendencies increases returns across games.
During continual play, each agent reflects after every game and summarizes past interactions when the context window fills.
As illustrated in \Cref{fig:teaser}a, we design a protocol that evaluates each agent with and without memory under identical conditions to isolate its \emph{experience gain} $\Delta_{\text{exp}}$ from single-game skill.

We benchmark open-weight and frontier language models under this protocol and find that memory brings little performance gain for open-weight models and inconsistent gains for frontier models.
To pinpoint where adaptation breaks down, we analyze the agents' written summaries alongside their gameplay.
\Cref{fig:teaser}b shows that models quickly identify and record opponent tendencies in memory, yet open-weight models rarely adjust their play and frontier models follow their stored plans inconsistently.
We name this disconnect between summarized observations in memory and downstream decisions the \emph{memory-doing gap}.

To close this gap, we construct a pool of exploitable opponents and post-train Qwen3.5-9B \citep{qwen2026qwen35} with supervised fine-tuning followed by trajectory-level reinforcement learning \citep{shao2024deepseekmath,wu2025resum,lu2025supo,yu2025memagent,li2026compactionrl}.
We optimize full-match return across context segments so summary generation and action selection are trained together.
On unseen opponents, our 9B RL policy doubles experience gain over supervised fine-tuning in Leduc Hold'em and increases it fivefold in No-Limit Hold'em, as previewed in \Cref{fig:teaser}c.
Across 35 opponents, our policy matches the strongest frontier model in Leduc Hold'em and surpasses all frontier models on experience gain and return in No-Limit Hold'em.
It also generalizes beyond its training horizon and retains most of its gain on summaries from frontier models, showing that RL teaches agents to act on stored observations.

Our main contributions are summarized as follows:
\begin{itemize}[leftmargin=1.4em,itemsep=2pt,topsep=2pt]
\item We design an evaluation protocol in repeated imperfect-information games that isolates \emph{experience gain} from single-game skill across continual interactions.
\item We benchmark open-weight and frontier language models and identify a \emph{memory-doing gap}, where models accurately record opponent tendencies in memory yet rarely depart from default play.
\item We develop a post-training recipe combining supervised fine-tuning and trajectory-level RL across context segments, enabling a 9B model to match or surpass frontier models on unseen opponents.
\end{itemize}

\section{Problem Formulation and Evaluation Protocol}
\label{sec:eval}

To study how language agents adapt their decisions across sequential episodes, \Cref{sec:eval_formulation} formalizes an agent with in-context memory and defines our metric for experience gain.
\Cref{sec:protocol} then introduces repeated imperfect-information games as a controlled testbed, and \Cref{sec:eval_protocol} describes our interaction and evaluation protocol.

\subsection{Continual Agents and Experience Gain}
\label{sec:eval_formulation}
\label{sec:eval_duplicate}

Consider a language model policy $\pi$ that interacts with an environment over $N$ sequential episodes indexed by $t \in \{1, \dots, N\}$.
Before episode $t$, the agent holds an in-context memory $M_{t-1}$ summarizing its past interactions, initialized at $M_0 = \varnothing$.
During episode $t$, the agent selects actions $a \sim \pi(\cdot \mid o, M_{t-1})$ conditioned on the current observation $o$ and its memory.
At the end of the episode, it receives a scalar reward and updates its memory to $M_t$.

An agent that learns from experience should make better decisions when given its memory than when acting without it.
Let $r_t(M_{t-1})$ denote the agent's reward in episode $t$ when retaining memory across episodes, and let $r_t(\varnothing)$ denote its reward when memory is reset to empty before every episode.
As shown in \Cref{fig:teaser}a, evaluating both runs under identical environment random seeds defines the \emph{experience gain} as the average per-episode improvement from memory,
\begin{equation}
\Delta_{\text{exp}} \;=\; \frac{1}{N}\sum_{t=1}^{N}\mathbb{E}\big[\,r_t(M_{t-1}) - r_t(\varnothing)\,\big].
\label{eq:exp_gain}
\end{equation}
Both runs start from an empty memory in the first episode and face identical environment realizations, so a positive $\Delta_{\text{exp}}$ isolates the reward gained from experience accumulated across episodes.

\subsection{Imperfect-Information Games as a Testbed}
\label{sec:eval_opponents}
\label{sec:eval_headroom}
\label{sec:results_setup}
\label{sec:protocol}

We instantiate this setting in repeated two-player zero-sum imperfect-information games.
Each episode is a stage game against a stationary opponent $\pi_{\text{opp}}$, and $N$ consecutive games form a match.
Our testbed covers Leduc Hold'em \citep{southey2005bayes}, a compact poker game with 936 information sets that permits exact analysis, and Heads-Up No-Limit Texas Hold'em \citep{moravcik2017deepstack,brown2018superhuman}, which features continuous bet sizing over four betting rounds, as detailed in \Cref{app:setting}.
Offline computation of a Nash equilibrium for a single game is now mature even in full-scale poker \citep{moravcik2017deepstack,brown2018superhuman,brown2019superhuman}.
Adapting online to an unknown opponent over repeated games remains an open challenge in game theory \citep{southey2005bayes,bard2013online}.
Repeated imperfect-information games offer two properties that make them well suited for evaluating in-context adaptivity.

\paragraph{Positive returns require adapting to the opponent.}
In standard single-agent benchmarks, a stronger model often solves more tasks on the first attempt without adapting across episodes.
In symmetric two-player zero-sum games, static single-game skill alone cannot achieve positive expected returns.
Let $u(\pi, \pi_{\text{opp}})$ denote the expected per-game reward of $\pi$ against $\pi_{\text{opp}}$, averaged over both starting seats.
A Nash equilibrium strategy $\pi^*$ guarantees $\min_{\pi_{\text{opp}}} u(\pi^*, \pi_{\text{opp}}) = 0$ \citep{kuhn1953extensive,shoham2008multiagent}.
Any fixed strategy earns zero against itself, so no memoryless policy can guarantee a positive expected reward against an unknown opponent.
Winning more against a flawed opponent requires inferring how it deviates from equilibrium and adjusting actions to exploit it \citep{billings1998opponent,southey2005bayes}.
The highest achievable expected reward is the opponent's exploitability $\operatorname{Expl}(\pi_{\text{opp}}) = \max_{\pi} u(\pi, \pi_{\text{opp}})$ \citep{johanson2007computing,lanctot2019openspiel}, which we compute exactly in Leduc Hold'em as described in \Cref{app:reward}.

\paragraph{Each game provides partial and noisy feedback.}
Unlike deterministic tasks where an agent can repeat a winning trajectory, a single poker hand reveals limited information about the opponent.
Private cards stay hidden whenever a player folds, and stochastic card deals create high reward variance even under optimal play.
Improving over a match therefore requires aggregating observations across games, separating opponent habits from card luck, and deciding when the evidence justifies departing from default play.

\subsection{Game Protocol and Context Compaction}
\label{sec:eval_protocol}
\label{sec:method_harness}

\paragraph{Stage-game interaction and solver-grounded reflection.}
Each stage game $t$ unfolds as a multi-turn interaction conditioned on the agent's current memory $M_{t-1}$.
At each turn, the agent observes the current game state and legal actions, reasons in natural language, and outputs an action.
The environment then returns the opponent's move and the next state.
When the game ends, the agent observes the hand outcome and writes a brief post-game reflection.
Human poker players routinely review completed hands with game-theoretic solvers to calibrate their decisions.
Following this practice, the reflection prompt includes reference equilibrium action probabilities from the solver for the completed hand, shown in \Cref{app:prompts} and ablated in \Cref{sec:results_q3}.
The completed game transcript and reflection remain in context as part of the agent's memory for subsequent games.

\paragraph{Context compaction.}
Keeping every raw transcript in context quickly exceeds a fixed context window of $L$ tokens over a multi-game match.
We therefore adopt the context-compaction loop used in long-horizon agent harnesses \citep{shinn2023reflexion,packer2023memgpt,anthropic2025claudecode,openai2025codex,wang2025openhands} and hold it fixed across all models.
The memory $M_{t-1} = (s, H)$ combines a natural-language summary $s$ consolidated from earlier games with a buffer $H$ of raw game transcripts and reflections accumulated since the last compaction.
Whenever $|s| + |H|$ approaches the window limit $L$ between games, the model compacts its context by generating an updated summary and resetting the buffer,
\begin{equation}
s' \;\sim\; \pi\big(\cdot \mid s, H\big), \qquad M_t \;\leftarrow\; \big(s', \varnothing\big),
\label{eq:compact}
\end{equation}
while retaining recent hand records alongside $s'$, as detailed in \Cref{app:compaction,app:notes}.

\paragraph{Paired evaluation.}
Each match is played twice on identical card deals, once retaining $M_{t-1}$ across games and once resetting $M_{t-1} = \varnothing$ before every game.
Every deal sequence is also replayed with the starting seats swapped, and the two mirrored matches are averaged into one paired sample to eliminate seat advantage and reduce card variance \citep{bowling2015heads,burch2018aivat}.
Alongside mean per-game \emph{return} and \emph{experience gain} $\Delta_{\text{exp}}$, we report match \emph{win rate}, defined as the fraction of matches with positive total reward where ties count as half a win.

\section{Evaluating In-Context Adaptivity in LLMs}\label{sec:stubborn}

\begin{table}[t]
\centering
\footnotesize
\setlength{\tabcolsep}{3.4pt}
\renewcommand{\arraystretch}{1.25}
\caption{\textbf{Memory brings small and inconsistent gains against RL and solver baselines.}
Each cell reports experience gain $\Delta_{\text{exp}}$ over $N = 50$ games, measured in chips per game for Leduc Hold'em and big blinds per game for No-Limit Hold'em.
Small grey numbers indicate 95\% confidence intervals, and bold values denote intervals that exclude zero.
The top row lists the exploitability $\operatorname{Expl}(\pi_{\text{opp}})$ of each Leduc Hold'em opponent.}
\label{tab:gain_models}
\resizebox{\textwidth}{!}{%
\begin{tabular}{l *{9}{c}}
\toprule
 & \multicolumn{6}{c}{\textbf{Leduc Hold'em} {\scriptsize chips / game}} & \multicolumn{3}{c}{\textbf{No-Limit Hold'em} {\scriptsize bb / game}} \\
\cmidrule(lr){2-7}\cmidrule(l){8-10}
 & CFR+ & Deep CFR & MMD & NFSP & DMC & DQN & NFSP & DMC & DQN \\
{\color{black!55}\textit{Exploitability} $\operatorname{Expl}$} & {\color{black!55}\textit{0.00}} & {\color{black!55}\textit{0.06}} & {\color{black!55}\textit{0.54}} & {\color{black!55}\textit{0.65}} & {\color{black!55}\textit{1.68}} & {\color{black!55}\textit{1.83}} & {\color{black!55}--} & {\color{black!55}--} & {\color{black!55}--} \\
\midrule
\rowcolor{tabPanel}\multicolumn{10}{c}{\textbf{\textit{Open-weight models}}} \\
\midrule
Qwen3.5-9B & \cellcolor{lossc!27}\textminus{}0.15\ci{0.25} & \cellcolor{gainc!11}+0.05\ci{0.24} & \cellcolor{lossc!11}\textminus{}0.05\ci{0.27} & \cellcolor{lossc!28}\textminus{}0.16\ci{0.21} & \cellcolor{lossc!18}\textminus{}0.09\ci{0.20} & \cellcolor{gainc!21}+0.11\ci{0.32} & \cellcolor{lossc!15}\textminus{}0.42\ci{1.11} & \cellcolor{lossc!8}\textminus{}0.17\ci{1.12} & \cellcolor{gainc!7}+0.12\ci{1.36} \\
Qwen3.5-27B & \cellcolor{lossc!26}\textbf{\textminus{}0.14}\ci{0.14} & \cellcolor{lossc!6}\textminus{}0.01\ci{0.24} & \cellcolor{lossc!21}\textminus{}0.11\ci{0.21} & \cellcolor{lossc!44}\textbf{\textminus{}0.26}\ci{0.14} & \cellcolor{gainc!25}+0.13\ci{0.17} & \cellcolor{gainc!37}+0.21\ci{0.25} & \cellcolor{lossc!6}\textminus{}0.06\ci{0.96} & \cellcolor{gainc!13}+0.36\ci{1.20} & \cellcolor{gainc!7}+0.13\ci{0.78} \\
Qwen3.5-122B & \cellcolor{lossc!17}\textminus{}0.08\ci{0.18} & \cellcolor{gainc!28}+0.15\ci{0.18} & \cellcolor{gainc!10}+0.04\ci{0.18} & \cellcolor{lossc!11}\textminus{}0.04\ci{0.24} & \cellcolor{lossc!19}\textminus{}0.10\ci{0.10} & \cellcolor{gainc!11}+0.04\ci{0.19} & \cellcolor{lossc!21}\textminus{}0.65\ci{1.29} & \cellcolor{gainc!15}+0.43\ci{0.59} & \cellcolor{gainc!26}\textbf{+0.85}\ci{0.80} \\
\midrule
\rowcolor{tabPanel}\multicolumn{10}{c}{\textbf{\textit{Proprietary models}}} \\
\midrule
Claude Opus 5.5 & \cellcolor{gainc!8}+0.03\ci{0.22} & \cellcolor{gainc!31}\textbf{+0.18}\ci{0.13} & \cellcolor{gainc!8}+0.02\ci{0.24} & \cellcolor{lossc!6}\textminus{}0.01\ci{0.17} & \cellcolor{gainc!31}+0.17\ci{0.18} & \cellcolor{gainc!50}\textbf{+0.30}\ci{0.19} & \cellcolor{lossc!26}\textminus{}0.87\ci{1.22} & \cellcolor{gainc!26}+0.85\ci{1.02} & \cellcolor{gainc!50}\textbf{+1.78}\ci{0.70} \\
Claude Sonnet 5 & \cellcolor{lossc!21}\textminus{}0.11\ci{0.18} & \cellcolor{gainc!31}\textbf{+0.17}\ci{0.14} & \cellcolor{gainc!7}+0.02\ci{0.12} & \cellcolor{lossc!30}\textbf{\textminus{}0.17}\ci{0.11} & \cellcolor{lossc!27}\textminus{}0.15\ci{0.22} & 0.00\ci{0.15} & \cellcolor{gainc!6}+0.07\ci{0.78} & \cellcolor{lossc!16}\textminus{}0.48\ci{0.70} & \cellcolor{lossc!9}\textminus{}0.18\ci{0.75} \\
Gemini 3.1 Pro & \cellcolor{gainc!7}+0.02\ci{0.12} & \cellcolor{gainc!15}+0.07\ci{0.07} & \cellcolor{lossc!6}\textminus{}0.01\ci{0.08} & \cellcolor{lossc!20}\textminus{}0.10\ci{0.16} & \cellcolor{lossc!13}\textminus{}0.06\ci{0.14} & \cellcolor{lossc!16}\textminus{}0.07\ci{0.11} & \cellcolor{gainc!10}+0.24\ci{0.48} & \cellcolor{gainc!12}+0.32\ci{0.52} & \cellcolor{gainc!6}+0.07\ci{0.58} \\
Gemini 3.7 Flash & \cellcolor{lossc!6}\textminus{}0.01\ci{0.12} & \cellcolor{gainc!30}\textbf{+0.17}\ci{0.10} & \cellcolor{gainc!43}\textbf{+0.25}\ci{0.11} & \cellcolor{lossc!13}\textminus{}0.06\ci{0.14} & \cellcolor{lossc!23}\textminus{}0.12\ci{0.14} & \cellcolor{gainc!43}\textbf{+0.25}\ci{0.21} & \cellcolor{lossc!11}\textminus{}0.29\ci{0.97} & \cellcolor{gainc!21}\textbf{+0.66}\ci{0.45} & \cellcolor{gainc!8}+0.17\ci{0.42} \\
\bottomrule
\end{tabular}%
}
\vspace{-0.4em}
\end{table}

\begin{figure}[t]
\centering
\includegraphics[width=\textwidth]{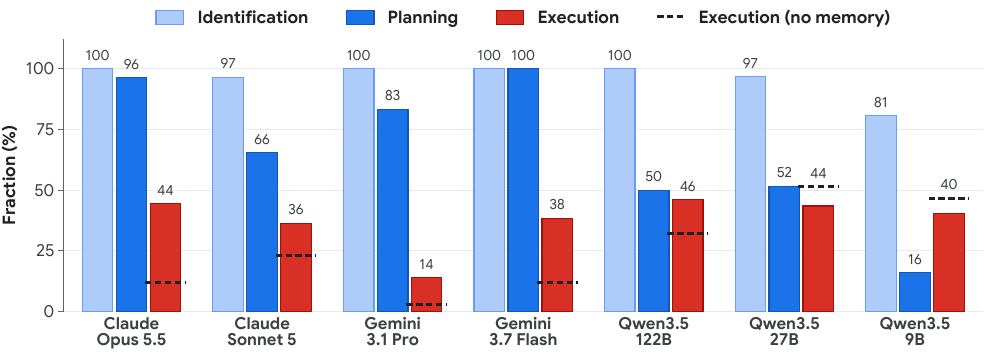}
\vspace{-1.2em}
\caption{\textbf{Models identify the opponent's weaknesses but rarely execute an exploit.}
Results in Leduc Hold'em against the four exploitable neural baselines.
Identification and planning are fractions of final summaries, and execution is a fraction of decisions, as defined in \Cref{sec:memory_gap}.
Dashed ticks mark execution in Game~1, when memory is empty and any switch comes from prior knowledge alone.}
\vspace{-0.5em}
\label{fig:memory_gap}
\end{figure}

We begin by evaluating how effectively current LLMs adapt from experience.
We measure their experience gain $\Delta_{\text{exp}}$ by playing against RL and solver baselines in poker, and then examine their written summaries and decisions to identify where adaptation breaks down.

\subsection{Benchmarking Experience Gain Against Baselines}

\paragraph{Setup.}
We evaluate the open-weight Qwen3.5 family at 9B, 27B, and 122B parameters \citep{qwen2026qwen35} alongside Claude Opus~5.5, Claude Sonnet~5, Gemini~3.1 Pro, and Gemini~3.7 Flash.
Each model plays $N = 50$-game matches with and without memory on identical card deals using the harness in \Cref{sec:method_harness}.
Opponents come from prior RL and solver poker evaluations \citep{guo2023suspicion,lin2026toolpoker}.
In Leduc Hold'em, we test against the equilibrium solver CFR+ \citep{tammelin2014solving} and five neural agents, including Deep CFR \citep{brown2019deep}, Magnetic Mirror Descent \citep{sokota2023unified}, Neural Fictitious Self-Play \citep{heinrich2016nfsp}, Deep Monte-Carlo \citep{zha2021douzero}, and DQN \citep{mnih2015human,zha2019rlcard}.
In No-Limit Hold'em, we evaluate against full-game NFSP, DMC, and DQN agents.

\paragraph{Memory brings small and inconsistent gains across model scales.}
\Cref{tab:gain_models} shows that no evaluated model improves reliably from memory.
Across all 63 model-opponent pairs, the 95\% confidence interval of $\Delta_{\text{exp}}$ excludes zero on the positive side in only 9 cases and on the negative side in 3.
The clearest gains appear against DMC and DQN, the two opponents with the most extreme flaws, where Claude Opus~5.5 gains $+0.30$ chips per game in Leduc Hold'em.
Against the more balanced NFSP agent, memory lowers the Leduc Hold'em return of every model.
Gains against CFR+ and Deep CFR, whose exploitability is near zero, reflect models correcting their own baseline mistakes rather than exploiting the opponent.
Scaling Qwen3.5 from 9B to 122B parameters strengthens play without memory yet produces no consistent experience gain.

\subsection{Diagnosing the Memory-Doing Gap}
\label{sec:memory_gap}

To understand why memory yields such small gains in \Cref{tab:gain_models}, we trace the steps required to adapt from experience.
Across a match, an agent must infer how the opponent plays from past hands, state a counter-strategy in its summary, and change its in-game actions to carry out that strategy.
A breakdown at any step prevents experience gain.
\Cref{fig:memory_gap} reports three metrics aligned with these steps.
\emph{Identification} is the fraction of matches whose final summary accurately states at least one exploitable weakness of the opponent.
\emph{Planning} is the fraction of matches whose summary also states a concrete plan to exploit that weakness.
\emph{Execution} is the fraction of decisions where memory switches the agent to the best-response action when its action without memory differs from the best response.

\paragraph{Evaluation setup.}
We evaluate all seven models from \Cref{tab:gain_models} in Leduc Hold'em against MMD, NFSP, DMC, and DQN, the four neural opponents that deviate from equilibrium.
For each opponent, we build a ground-truth behavioral profile from its action frequencies across the game tree.
An LLM judge scores \emph{Identification} and \emph{Planning} by comparing the final summary $s$ of each match against the opponent's profile without seeing the model identity.
To score \emph{Execution}, we use the exact best-response policy in Leduc Hold'em on shared decision states in games 11 to 50 and compare it with the same rate in Game~1, when memory is still empty.
\Cref{app:audit} details profile construction, blind judging, and state filtering.

\paragraph{Models identify weaknesses and form plans in memory, but rarely execute them.}
\Cref{fig:memory_gap} reveals a sharp drop between summarizing the opponent and acting on that summary.
Averaged across all seven models, \emph{Identification} reaches $96\%$ and \emph{Planning} reaches $65\%$, confirming that current models reliably detect and record how their opponent plays.
Yet \emph{Execution} averages only $37\%$, compared with $25\%$ in Game~1, when memory is still empty and switches come from action sampling alone.
Frontier models rise well above their Game~1 rate, led by Claude Opus~5.5, whereas the smaller Qwen3.5 models stay at or below it.

\paragraph{Followed plans rarely change default actions in No-Limit Hold'em.}
When a game is too large to compute an exact best response, we test whether a model's decisions follow the plan written in its own summary.
In No-Limit Hold'em, we sample 30 decision states per model against NFSP, DMC, and DQN across six models, as described in \Cref{app:audit}.
At each state, the blind judge reads the active summary $s$ and the game state to check whether the summary recommends a specific action.
The summary gives a clear action recommendation in only 38 of the 180 sampled states, staying too general to guide a concrete choice in the rest.
The model follows that recommendation in 33 of those 38 states, but in 27 of them it already takes the exact same action without memory.
Even when models follow their notes in No-Limit Hold'em, their notes mostly recommend actions they already choose by default.

\paragraph{The memory-doing gap.}
Prior work documents a knowing-doing gap in single-episode tasks such as multi-armed bandits, tic-tac-toe, and single-hand poker, where models fail to act on knowledge they can state explicitly \citep{schmied2025greedy,lin2026toolpoker}.
Our results reveal an analogous \emph{memory-doing gap} in continual interaction, where the unused knowledge is inferred online from past episodes and stored in memory.
In Leduc Hold'em, models write an exploit plan yet rarely switch their actions to execute it.
In No-Limit Hold'em, their summaries rarely prescribe actions that depart from single-game play.
Prior work on agent memory focuses on how experience is stored and retrieved \citep{shinn2023reflexion,packer2023memgpt}.
Our findings show that models already record the opponent accurately and instead struggle to act on that record.

\section{Post-Training for Continual In-Context Adaptation}
\label{sec:method}

\begin{figure*}[t]
\centering
\includegraphics[width=\textwidth]{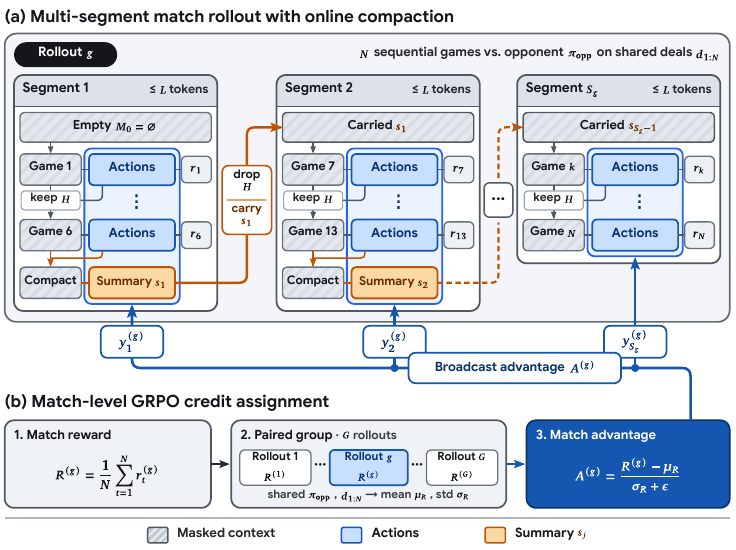}
\vspace{-1.2em}
\caption{\textbf{Trajectory-level RL across compacted matches.}
\textbf{(a)}~In rollout $g$, each segment accumulates game history $H$ up to $L$ tokens, then drops $H$ and carries the summary $s_j$ into the next segment as masked context.
\textbf{(b)}~Across $G$ paired rollouts on shared card deals $d_{1:N}$, the standardized match advantage $A^{(g)}$ is broadcast to all generated tokens $y^{(g)}_j$ across segments while context tokens are masked.}
\vspace{-0.6em}
\label{fig:agent}
\end{figure*}

We next study how post-training can close the memory-doing gap.
\Cref{sec:method_opponents} builds a pool of exploitable opponents, and \Cref{sec:method_sft,sec:method_rl} train a 9B policy via supervised fine-tuning followed by trajectory-level reinforcement learning.

\subsection{Constructing the Opponent Pool}
\label{sec:method_opponents}

Training an agent to infer and exploit opponent tendencies requires a broad population of suboptimal opponents.
Existing poker benchmarks offer only a handful of fixed opponents per game, and most approximate a Nash equilibrium rather than displaying varied flaws.
We therefore build a pool of code-based opponent policies from two sources detailed in \Cref{app:opponents}.
First, we design heuristic policies for common archetypes such as folding too often to raises, calling nearly every bet, or sizing bets by hand strength.
Second, we evolve policies with Code-Space Response Oracles \citep{alphaevolve2026}, keeping checkpoints with distinct betting styles against the expanding population.
We split the pool into seen training opponents and unseen evaluation opponents, and train separate models for Leduc Hold'em and No-Limit Hold'em.

\subsection{Supervised Fine-Tuning with Context Compaction}
\label{sec:method_sft}

Before an agent can adapt across a match, it must play individual games competently and write structured summaries.
Off-the-shelf Qwen3.5-9B plays weakly without memory and rarely turns its observations into concrete plans, as shown in \Cref{tab:gain_models,fig:memory_gap}.
We warm-start Qwen3.5-9B on 50-game matches played by Gemini~3.7 Flash against training opponents under the harness of \Cref{sec:method_harness}.
Full matches exceed the context window limit $L$, so we split each recorded match at its compaction boundaries into standalone segments.
As shown in \Cref{fig:agent}a, segment $j$ starts from the carried summary $s_{j-1}$, collects game transcripts and reflections in $H$, and ends with the updated summary $s_j$.
We train with cross-entropy loss on the teacher's reasoning, actions, reflections, and summaries while masking clear decision errors as described in \Cref{app:training}.
Imitating isolated segments still inherits the teacher's weak adaptation across games, motivating reinforcement learning over complete matches.

\subsection{Trajectory-Level GRPO Across Compacted Matches}
\label{sec:method_rl}

An observation recorded after game $t$ affects rewards only in later games $t+1, \dots, N$, often after one or more context compactions.
We therefore roll out complete $N$-game matches with online compaction during RL and assign one match-level advantage across every segment of the match \citep{wu2025resum,lu2025supo,yu2025memagent,li2026compactionrl}.

\paragraph{Paired group rollouts and match reward.}
Each RL step samples a batch of rollout groups in parallel.
For each group, we draw one training opponent $\pi_{\text{opp}}$ from the pool and one shared sequence of $N = 20$ card deals $d_{1:N} = (d_1, \dots, d_N)$.
We then roll out $G$ complete matches independently under $\pi_\theta$ against that same opponent $\pi_{\text{opp}}$ on the identical deal sequence $d_{1:N}$, rotating starting seats every game.
Sharing both the opponent and the card deals across all $G$ rollouts cancels card luck within each group.
Reward differences inside a group therefore reflect how well the policy gathers and exploits opponent tendencies rather than which hands were dealt.
Rollout $g \in \{1, \dots, G\}$ spans $S_g$ compacted segments linked by summaries $s_1, \dots, s_{S_g - 1}$ and collects per-game rewards $r_1^{(g)}, \dots, r_N^{(g)}$.
The match reward is simply the average per-game reward across the match,
\begin{equation}
R^{(g)} \;=\; \frac{1}{N}\sum_{t=1}^{N} r_t^{(g)}.
\label{eq:norm_return}
\end{equation}

\paragraph{Advantage broadcast across segments and policy update.}
Within each group, we standardize the match rewards $\{R^{(g)}\}_{g=1}^{G}$ into the group-relative advantage \citep{shao2024deepseekmath},
\begin{equation}
A^{(g)} \;=\; \frac{R^{(g)} - \mu_R}{\sigma_R + \epsilon},
\label{eq:match_adv}
\end{equation}
with group mean $\mu_R$, group standard deviation $\sigma_R$, and $\epsilon = 10^{-6}$.
Per-group standardization puts groups facing differently exploitable opponents on a common unit-variance scale without manual reward shaping.
As shown in \Cref{fig:agent}b, we unpack rollout $g$ into its $S_g$ context segments, each of at most $L$ tokens.
In segment $j \in \{1, \dots, S_g\}$, context tokens comprising $s_{j-1}$ and game observations are masked from the loss.
Every generated token $y^{(g)}_{j,k}$ in segment $j$, spanning reasoning, actions, reflections, and the summary $s_j$, receives the match-level advantage $A^{(g)}$.
We maximize the clipped GRPO objective without a KL penalty \citep{shao2024deepseekmath,yu2025dapo}.
For one group, the objective is
\begin{equation}
\begin{aligned}
\mathcal{J}(\theta) \;=\; {}& \frac{1}{\sum_{g=1}^{G}\sum_{j=1}^{S_g}\big|y^{(g)}_j\big|}\,\sum_{g=1}^{G}\sum_{j=1}^{S_g}\sum_{k=1}^{|y^{(g)}_j|}
\min\Bigg(\frac{\pi_\theta\big(y^{(g)}_{j,k} \mid h^{(g)}_{j,k}\big)}{\pi_{\theta_{\text{old}}}\big(y^{(g)}_{j,k} \mid h^{(g)}_{j,k}\big)}\, A^{(g)}, \\[6pt]
& \qquad\qquad \operatorname{clip}\bigg(\frac{\pi_\theta\big(y^{(g)}_{j,k} \mid h^{(g)}_{j,k}\big)}{\pi_{\theta_{\text{old}}}\big(y^{(g)}_{j,k} \mid h^{(g)}_{j,k}\big)},\; 1-\varepsilon_{\text{low}},\; 1+\varepsilon_{\text{high}}\bigg)\, A^{(g)}\Bigg),
\end{aligned}
\label{eq:grpo}
\end{equation}
where $h^{(g)}_{j,k}$ denotes all preceding tokens in segment $j$, with $\varepsilon_{\text{low}} = 0.2$ and $\varepsilon_{\text{high}} = 0.28$.
Dividing by total generated tokens weights every rollout group equally, while single-window segments keep memory cost flat and credit early summaries for later rewards.

\section{Experiments and Analysis}
\label{sec:results}

We now evaluate whether post-training across full matches teaches language models to act on their stored observations over long horizons.
We outline the setup in \Cref{sec:results_setup_main}, present the main results in \Cref{sec:main-results}, analyze the memory-doing gap in \Cref{sec:rl-teaches,sec:case_study}, and ablate our design choices in \Cref{sec:ablations}.

\subsection{Experimental Setup}
\label{sec:results_setup_main}

\begin{table}[!t]
\centering
\footnotesize
\setlength{\tabcolsep}{3.1pt}
\renewcommand{\arraystretch}{1.18}
\caption{\textbf{Continual evaluation across 35 opponents over 50-game matches.}
Columns report mean return per game with memory, paired experience gain $\Delta_{\text{exp}}$, and match win rate on unseen opponents.
Best model per column in bold, second-best underlined.}
\label{tab:main}
\label{tab:headline}
\resizebox{\textwidth}{!}{%
\begin{tabular}{l rrrrr rrrrr}
\toprule
 & \multicolumn{5}{c}{\textbf{Leduc Hold'em} {\scriptsize chips / game}} & \multicolumn{5}{c}{\textbf{No-Limit Hold'em} {\scriptsize bb / game}} \\
\cmidrule(lr){2-6}\cmidrule(l){7-11}
 & \multicolumn{3}{c}{Return with memory} & \multicolumn{2}{c}{Unseen {\scriptsize(7)}} & \multicolumn{3}{c}{Return with memory} & \multicolumn{2}{c}{Unseen {\scriptsize(7)}} \\
\cmidrule(lr){2-4}\cmidrule(lr){5-6}\cmidrule(lr){7-9}\cmidrule(l){10-11}
\textbf{Model} & RL \& Solver {\scriptsize(6)} & Seen {\scriptsize(6)} & Unseen {\scriptsize(7)} & $\Delta_{\text{exp}}$ & Win \% & RL \& Solver {\scriptsize(5)} & Seen {\scriptsize(4)} & Unseen {\scriptsize(7)} & $\Delta_{\text{exp}}$ & Win \% \\
\midrule
Claude Opus 5.5 & \textbf{+0.28} & \underline{+0.91} & \textbf{+0.45} & \textbf{+0.25} & \textbf{76\%} & \textbf{+0.64} & \textbf{+1.06} & \underline{+1.02} & +0.19 & \underline{65\%} \\
Claude Sonnet 5 & +0.13 & +0.63 & +0.25 & +0.07 & 63\% & \textminus{}0.28 & +0.89 & +0.69 & +0.11 & 59\% \\
Gemini 3.1 Pro & +0.12 & +0.56 & +0.24 & +0.06 & 70\% & \textminus{}0.16 & +0.24 & +0.20 & \underline{+0.29} & 50\% \\
Gemini 3.7 Flash & \underline{+0.21} & +0.80 & +0.30 & +0.12 & 70\% & \underline{+0.42} & +0.22 & +0.31 & \textminus{}0.03 & 52\% \\
\cmidrule(l{2pt}r{2pt}){1-11}
Qwen3.5-122B & +0.06 & +0.52 & +0.14 & +0.09 & 56\% & \textminus{}0.44 & +0.59 & +0.07 & +0.07 & 52\% \\
Qwen3.5-27B & +0.01 & +0.50 & +0.15 & +0.12 & 58\% & \textminus{}0.62 & +0.38 & \textminus{}0.10 & +0.22 & 46\% \\
Qwen3.5-9B & \textminus{}0.19 & +0.19 & \textminus{}0.07 & 0.00 & 46\% & \textminus{}0.51 & \textminus{}0.63 & \textminus{}0.51 & +0.06 & 37\% \\
\hspace{1.25em}+\,SFT & +0.19 & +0.76 & +0.29 & +0.11 & 67\% & +0.27 & +0.36 & +0.42 & +0.06 & 57\% \\
\rowcolor{tabRL}\hspace{1.25em}+\,\textbf{RL} & +0.16 & \textbf{+0.95} & \underline{+0.40} & \underline{+0.23} & \underline{71\%} & \underline{+0.42} & \underline{+1.03} & \textbf{+1.28} & \textbf{+0.30} & \textbf{66\%} \\
\bottomrule
\end{tabular}%
}
\vspace{-0.4em}
\end{table}

\paragraph{Models and training.}
We post-train Qwen3.5-9B~\citep{qwen2026qwen35} separately for each game, first with supervised fine-tuning to obtain \sftm{} and then with trajectory-level RL to obtain \ours{}.
Each RL rollout is a 20-game match against a training opponent, using group sizes of $G = 12$ in Leduc Hold'em and $G = 8$ in No-Limit Hold'em on shared card deals.
We compare against four frontier models and off-the-shelf Qwen3.5 at 9B, 27B, and 122B parameters.
Every model uses the compaction harness of \Cref{sec:method_harness} with a $32$K-token window. \Cref{app:training} gives full training hyperparameters.

\paragraph{Evaluation suite and match budget.}
We evaluate all models over $N = 50$-game matches across 35 opponents, with 19 in Leduc Hold'em and 16 in No-Limit Hold'em, plus a No-Limit MCCFR blueprint for our post-trained models in \Cref{app:curves}.
\textbf{RL and solver baselines} cover CFR+ and five neural agents in Leduc Hold'em and the five neural agents in No-Limit Hold'em.
\textbf{Seen opponents} are code-based policies from the training pool tested on fresh deals, and \textbf{Unseen opponents} are seven held-out code-based policies per game.
Our post-trained checkpoints play 128 paired matches per opponent.
Off-the-shelf models play up to 128 matches per opponent without memory and 16 matches in Leduc Hold'em or up to 32 in No-Limit Hold'em with memory due to proprietary API costs.

\begin{figure}[!t]
\centering
\includegraphics[width=\textwidth]{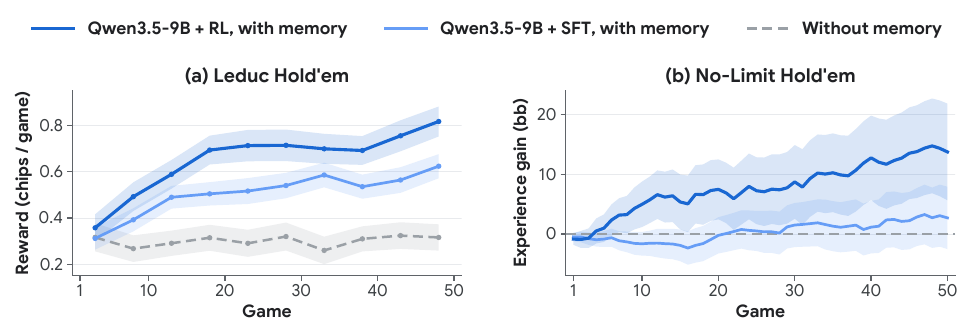}
\vspace{-1.4em}
\caption{\textbf{Continual adaptation across 50-game matches.}
\textbf{(a)}~Per-game reward in Leduc Hold'em and \textbf{(b)}~cumulative experience gain in No-Limit Hold'em across seen and unseen code-based opponents, with 95\% confidence bands.}
\label{fig:curve}
\vspace{-0.5em}
\end{figure}

\subsection{Main Results}
\label{sec:main-results}
\label{sec:results_q1}
\label{sec:curve}
\label{sec:results_q2}

\paragraph{\texorpdfstring{\ours{}}{Qwen3.5-9B + RL} matches or outperforms frontier models.}
\Cref{tab:main} compares our post-trained checkpoints against four frontier models and the off-the-shelf Qwen3.5 family.
Off-the-shelf open-weight models earn low returns with memory and gain little on unseen opponents even at 122B parameters.
Across the ten columns of \Cref{tab:main}, \ours{} places first or second in nine.
In Leduc Hold'em, \ours{} achieves the highest return on seen opponents and reaches +0.40 chips per game on unseen opponents, second only to Claude Opus~5.5.
In No-Limit Hold'em, \ours{} ranks second behind Claude Opus~5.5 on the RL and solver baselines and leads all models on unseen opponents at +1.28~bb per game.

\begin{figure}[!t]
\centering
\begin{minipage}[b]{0.63\textwidth}
\centering
\includegraphics[height=2.3in]{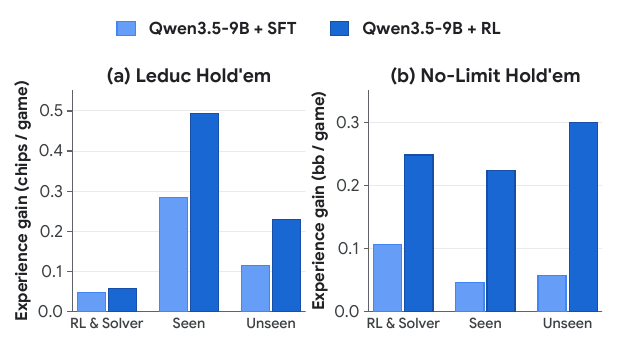}
\end{minipage}\hfill
\begin{minipage}[b]{0.345\textwidth}
\centering
\includegraphics[height=2.3in]{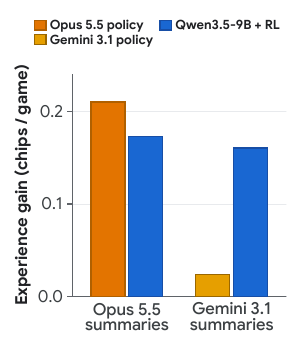}
\end{minipage}
\vspace{-0.6em}

\begin{minipage}[t]{0.63\textwidth}
\caption{\textbf{Experience gain $\Delta_{\text{exp}}$ from trajectory-level RL.}
Experience gain of \sftm{} and \ours{} against each opponent group in \textbf{(a)}~Leduc Hold'em and \textbf{(b)}~No-Limit Hold'em, with 95\% intervals in \Cref{tab:gain_groups}.}
\label{fig:gain}
\end{minipage}\hfill
\begin{minipage}[t]{0.345\textwidth}
\caption{\textbf{Cross-model summary transfer.}
Experience gain when each decision policy acts on a compaction model's summaries.}
\label{fig:swap}
\end{minipage}
\vspace{-0.5em}
\end{figure}

\paragraph{Trajectory-level RL increases experience gain over SFT in both games.}
Comparing the RL and SFT checkpoints isolates what trajectory-level RL adds beyond behavior cloning.
In Leduc Hold'em, both checkpoints earn similar returns without memory in \Cref{tab:nomem}, yet RL doubles unseen experience gain from $+0.11$ to $+0.23$ chips per game in \Cref{fig:gain}a.
In No-Limit Hold'em, \sftm{} inherits its teacher's weak adaptation on unseen opponents.
Trajectory-level RL strengthens play without memory in \Cref{tab:nomem} and lifts unseen experience gain fivefold to $+0.30$~bb per game in \Cref{fig:gain}b, leading all models.

\paragraph{In-context adaptation generalizes beyond the training horizon.}
RL rollouts last only 20 games during training, so we test whether the learned adaptation holds over 50-game evaluation matches.
\Cref{fig:curve} plots per-game reward in Leduc Hold'em and cumulative experience gain in No-Limit Hold'em across the 50 games of a match.
Without memory, rewards remain flat throughout the match.
With memory retained, \ours{} starts at the same level in Game~1, improves over the first 25 games, and sustains higher rewards through Game~50.

\subsection{Analysis of the Memory-Doing Gap}
\label{sec:rl-teaches}
\label{sec:drivers}
\label{sec:results_mechanisms}

\paragraph{Trajectory-level RL closes the execution gap of the SFT baseline.}
To pinpoint how RL improves adaptation, we evaluate the SFT and RL checkpoints on the identification, planning, and execution metrics of \Cref{sec:memory_gap} against the four exploitable neural opponents in Leduc Hold'em.
Both checkpoints identify the opponent's weakness in nearly every match and state a concrete exploit plan in most final summaries.
However, trajectory-level RL raises in-game execution from $34\%$ for \sftm{} to $44\%$ for \ours{}, matching Claude Opus~5.5 on the same opponents.
Rewarding full-match return across compacted segments trains the policy to execute the plans in its summary rather than forming sharper card beliefs, as a side probe in \Cref{app:probe} confirms.

\paragraph{Reinforcement learning translates summary observations into decisive actions.}
A similar shift appears in No-Limit Hold'em against an unseen opponent that folds too often to bets.
The SFT model, the RL policy, and Claude Opus~5.5 all record this habit in most summaries, yet \sftm{} and Claude Opus~5.5 barely raise their bluff frequency with weak hands.
For \ours{}, memory raises the bluff rate by 19 percentage points to $61\%$.
Across all unseen No-Limit Hold'em opponents, memory earns $+15$~bb of extra winnings per match for \ours{} versus $+3$~bb for \sftm{} in \Cref{fig:teaser}c.

\subsection{Case Study}
\label{sec:case_study}

\begin{figure}[t]
\centering
\includegraphics[width=\textwidth]{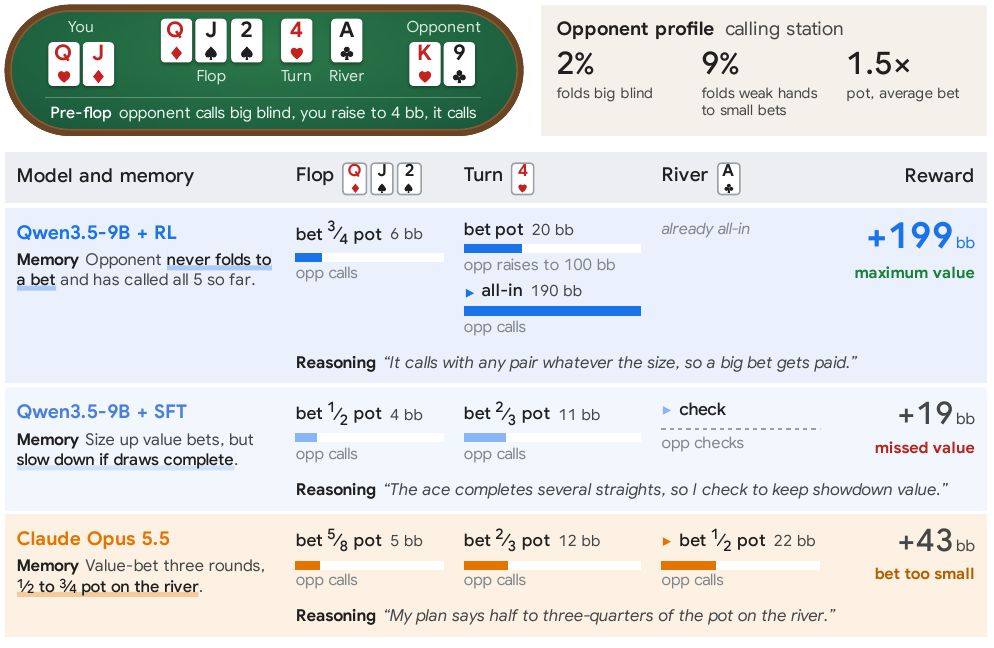}
\vspace{-1.4em}
\caption{\textbf{Qualitative comparison against an unseen calling-station opponent in No-Limit Hold'em.}
The top row shows the shared cards and the opponent's profile.
Each model row displays its memory before the hand and its actions in each betting round, with bars scaled to the bet size in big blinds and opponent replies in grey.
Triangles mark the key decision whose reasoning is quoted alongside the final hand reward.}
\vspace{-0.5em}
\label{fig:showcase_station}
\end{figure}

\paragraph{The RL policy acts on its summary to extract maximum value.}
\Cref{fig:showcase_station} compares our RL and SFT checkpoints with Claude Opus~5.5 on the same No-Limit Hold'em deal against an unseen opponent that calls nearly every bet.
Before this hand, all three models have recorded in memory that the opponent rarely folds, and each flops top two pair with queen-jack.
Both baselines hedge their stored plans with standard caution and bet only half to two-thirds of the pot on the flop and turn.
When an ace arrives on the river, \sftm{} checks out of fear of completed straights to win $+19$~bb, and Claude Opus~5.5 bets half the pot to win $+43$~bb.
Instead, \ours{} notes five prior calls, bets large on the flop and turn, and moves all-in over the opponent's turn raise to win $+199$~bb.

\subsection{Ablation Studies}
\label{sec:ablations}
\label{sec:results_q3}

\paragraph{The RL policy transfers to summaries written by frontier models.}
Trajectory-level RL trains compaction and action selection together.
We therefore test whether \ours{} relies on its own note format or acts on general natural-language summaries.
At each compaction step across nine Leduc Hold'em opponents, we replace the model's own summary with the summary generated by Claude Opus~5.5 or Gemini~3.1 Pro.
\Cref{fig:swap} shows that \ours{} retains 81\% of its own experience gain when given summaries from Claude Opus~5.5 and 75\% when given summaries from Gemini~3.1 Pro.
Pairing Gemini~3.1 Pro summaries with the RL actor lifts experience gain eightfold to $+0.16$ chips per game.
Frontier summaries already contain actionable observations that frozen policies leave unused.

\begin{table}[t]
\centering
\footnotesize
\setlength{\tabcolsep}{10pt}
\renewcommand{\arraystretch}{1.15}
\caption{\textbf{Ablating test-time solver feedback for \ours{}.}
Mean per-game return with memory across unseen opponents.}
\label{tab:solver_main}
\begin{tabular}{l rr}
\toprule
\textbf{Test-time setting} & \textbf{Leduc Hold'em} {\scriptsize(chips / game)} & \textbf{No-Limit Hold'em} {\scriptsize(bb / game)} \\
\midrule
With solver feedback & +0.47 & +1.07 \\
Without solver feedback & +0.49 & +0.97 \\
\midrule
Difference (95\% CI) & +0.02{\scriptsize\,\textpm{}0.07} & \textminus{}0.10{\scriptsize\,\textpm{}0.53} \\
\bottomrule
\end{tabular}
\vspace{-0.4em}
\end{table}

\paragraph{Test-time adaptation does not rely on external solver feedback.}
By default, the post-game reflection prompt includes reference equilibrium action probabilities from a game-theoretic solver for the completed hand.
We ablate this input at test time by evaluating \ours{} without solver feedback across six unseen opponents in Leduc Hold'em and five in No-Limit Hold'em over 32 paired 50-game matches per opponent.
\Cref{tab:solver_main} shows that returns with and without test-time solver feedback differ by at most $0.06$ units and lie well within the 95\% confidence intervals in both games.
The RL policy therefore infers and exploits opponent tendencies directly from game history at test time without requiring a solver.

\FloatBarrier

\section{Related Work}
\label{sec:related}

\paragraph{Agent memory and long-horizon harnesses.}
Many agent architectures introduce external memory structures to retain information over long horizons.
Early systems augment language models with verbal reflection, retrieval stores, or hierarchical context managers \citep{shinn2023reflexion,park2023generative,wang2023voyager,packer2023memgpt}.
Modern coding harnesses instead rely on periodic context compaction \citep{anthropic2025claudecode,openai2025codex,wang2025openhands}.
Recent work applies reinforcement learning to train the summarization or memory-curation module itself \citep{zhou2025mem1,yu2025memagent,wu2025resum,lu2025supo,cai2026player,li2026compactionrl}.
For example, MemoPilot trains a separate memory copilot over five-game sessions while keeping the decision policy frozen \citep{cai2026player}.
Despite these advances, multi-session benchmarks find that language models struggle to improve decisions from stored experience \citep{he2026memoryarena,zhao2026amabench,singh2026plasticity,shah2026gitswarm}.
We hold a standard reflection-and-compaction harness fixed over 50-game matches and study why the decision policy fails to act on its stored observations.

\paragraph{Reinforcement learning for decision-making agents.}
\citet{schmied2025greedy} show a knowing-doing gap in multi-armed bandits and tic-tac-toe, where models fail to take actions implied by their static knowledge until trained with RL.
We identify a parallel memory-doing gap in continual interaction, where the unused knowledge is inferred online from past episodes and stored in compacted summaries.
Cross-episode meta-RL methods train agents to explore and exploit across repeated attempts \citep{jiang2025lamer,lin2026orbit,yang2026mage}, assuming the entire interaction history fits inside one context window.
Our 50-game matches exceed a single context window.
We therefore roll out trajectories with online compaction and broadcast match-level credit across segments to optimize summary writing and action selection together.

\paragraph{LLMs and opponent modeling in imperfect-information games.}
Imperfect-information games are a classic testbed for strategic reasoning under uncertainty, spanning matrix games \citep{akata2023playing}, multi-game suites \citep{lanctot2019openspiel,duan2024gtbench}, and poker \citep{guo2023suspicion,zhuang2025pokerbench,lin2026toolpoker,xia2026iso}.
Single-hand LLM poker studies either imitate solver actions \citep{zhuang2025pokerbench} or call external solvers at decision time \citep{lin2026toolpoker}.
Multi-hand studies prompt frozen models to track opponent beliefs \citep{guo2023suspicion}, train a separate memory copilot \citep{cai2026player}, or train with a learned strategic reward model \citep{xia2026iso}.
We instead isolate adaptation with a paired control without memory and train one policy end-to-end on match reward across compacted context.
In computer poker, counterfactual regret minimization and deep self-play approximate Nash equilibria that guarantee zero expected reward \citep{zinkevich2007regret,lanctot2009monte,tammelin2014solving,bowling2015heads,moravcik2017deepstack,brown2018superhuman,brown2019deep}.
Winning more against flawed opponents requires opponent modeling or online best-response search \citep{billings1998opponent,southey2005bayes,johanson2007computing,ganzfried2011game,he2016opponent,rabinowitz2018machine}.
Rather than building game-specific opponent models or querying solvers during play, we train a language model policy to summarize opponent habits and adjust its actions in natural language.

\FloatBarrier
\section{Conclusion}
\label{sec:conclusion}

We study how LLM agents adapt their decisions from experience across repeated imperfect-information games.
Comparing runs with and without memory under identical conditions isolates the performance gained from accumulated experience.
Across Leduc Hold'em and No-Limit Hold'em, memory brings little performance gain for open-weight models and yields uneven improvements for frontier models.
Tracing summaries and decisions reveals a memory-doing gap where models accurately record opponent tendencies yet rarely depart from default single-game play.
Post-training a 9B model across compacted matches reaches frontier-level experience gain on unseen opponents and transfers to summaries from frontier models.
Together, these results show that continual adaptation hinges on training the policy to act on stored observations rather than redesigning the memory store.

\paragraph{Limitations and future work.}
Our study holds the opponent policy stationary within each match to isolate single-agent adaptation.
Future work can extend to co-adapting multi-agent settings.
Applying the same training recipe to multi-session coding and tool-use tasks where agents refine behavior across compacted windows is another natural direction.

\bibliography{main}

\appendix

\section{Implementation Details}
\label{app:harness}

\subsection{Poker Environments}
\label{app:setting}

\paragraph{Leduc Hold'em.}
\label{app:reward}
We build both poker environments on top of the OpenSpiel game engine \citep{lanctot2019openspiel} to evaluate adaptation across game trees that differ in size by orders of magnitude.
Leduc Hold'em \citep{southey2005bayes} is a compact poker variant played with a six-card deck of two suits of Jack, Queen, and King across two betting rounds.
Each player posts a $1$-chip ante before the deal, and each round allows at most two raises with fixed bet sizes of $2$ chips in the first round and $4$ chips in the second.
We report rewards in chips per game.
Every stationary opponent defines fixed action probabilities at each information set, allowing us to compute its exact best response and exploitability $\operatorname{Expl}(\pi_{\text{opp}}) = \max_{\pi} u(\pi, \pi_{\text{opp}})$ by full-tree traversal.

\paragraph{No-Limit Hold'em.}
Heads-Up No-Limit Texas Hold'em is the standard full-scale benchmark for imperfect-information games \citep{moravcik2017deepstack,brown2018superhuman}, played with a 52-card deck across four betting rounds.
Each player starts every game with a reset stack of $20{,}000$ chips under blinds of $50$ and $100$ chips, corresponding to $200$ big blinds.
At each turn, the agent may choose any bet or raise size in chips.
Sizes below the legal minimum are rounded up to that minimum, and sizes exceeding the remaining stack are treated as all-in.
We report bet sizes and rewards in big blinds per game, where one big blind equals $100$ chips.

\subsection{Opponent Pool Construction}
\label{app:opponents}

\paragraph{Design principle.}
Training an adaptive policy requires opponents with opposing weaknesses so that no single fixed strategy dominates the pool.
We therefore balance each training pool between opponents that fold too often and opponents that call too often.
Neither constant bluffing nor passive value betting beats the equilibrium policy across the pool, so the model can improve only by inferring how each opponent plays.
Code-based opponents come from two sources combining hand-engineered rules with Python programs evolved via Code-Space Response Oracles (CSRO) \citep{alphaevolve2026}.
\Cref{tab:opp_pool} summarizes how each source is split across supervised fine-tuning, RL training, and evaluation.
Evaluation further holds out an equilibrium solver and five neural policies per game from Deep CFR \citep{brown2019deep}, Magnetic Mirror Descent \citep{sokota2023unified}, Neural Fictitious Self-Play \citep{heinrich2016nfsp}, Deep Monte-Carlo \citep{zha2021douzero}, and DQN \citep{mnih2015human,zha2019rlcard}.

\begin{table}[h]
\centering
\footnotesize
\setlength{\tabcolsep}{6pt}
\renewcommand{\arraystretch}{1.12}
\caption{\textbf{Code-based opponents by construction source.}
Counts of opponents in the supervised fine-tuning data, the RL training pool, and the evaluation suite.
Seen opponents appear either during SFT or RL, while unseen opponents are held out from both training stages.}
\label{tab:opp_pool}
\begin{tabular}{l cccc cccc}
\toprule
 & \multicolumn{4}{c}{\textbf{Leduc Hold'em}} & \multicolumn{4}{c}{\textbf{No-Limit Hold'em}} \\
\cmidrule(lr){2-5}\cmidrule(l){6-9}
 & & & \multicolumn{2}{c}{Evaluation} & & & \multicolumn{2}{c}{Evaluation} \\
\cmidrule(lr){4-5}\cmidrule(l){8-9}
\textbf{Source} & SFT & RL & Seen & Unseen & SFT & RL & Seen & Unseen \\
\midrule
Hand-engineered & 20 & 26 & 4 & 4 & 15 & 11 & 3 & 4 \\
CSRO-evolved & 6 & 6 & 2 & 3 & 17 & 14 & 1 & 3 \\
\midrule
\textbf{Total} & \textbf{26} & \textbf{32} & \textbf{6} & \textbf{7} & \textbf{32} & \textbf{25} & \textbf{4} & \textbf{7} \\
\bottomrule
\end{tabular}
\end{table}

\paragraph{Hand-engineered opponents.}
In Leduc Hold'em, we construct heuristic opponents by applying scoped edits to the CFR+ equilibrium $\pi^*$ across selected rounds, seats, card ranks, and betting histories.
Each edit blends $\pi^*$ at severity $\alpha \in (0, 1]$ toward a target behavior such as folding weak hands, calling, raising, suppressing bluffs, or slow-playing.
Full-scale No-Limit Hold'em lacks a tabular equilibrium to perturb directly.
We instead define rule-based personas over five pre-flop tiers and 900 post-flop cells conditioned on round, hand strength, board texture, faced bet size, and seat, with additional rules for draws and board changes.
We add a candidate persona only if its action frequencies on probe states differ by at least $0.15$ from every existing opponent in the pool.

\paragraph{CSRO-evolved programs.}
Hand-engineered rules only capture failure modes specified in advance.
To generate broader strategic weaknesses, we use CSRO \citep{alphaevolve2026} to evolve standalone Python opponent programs as approximate best responses against an expanding population.
Each program implements its own hand evaluation, equity estimation, and bet sizing without sharing solver abstractions.
Retaining checkpoints from different stages of population training yields opponents with varied exploitable habits.
For example, evolved Leduc Hold'em bots build check-raise traps with strong hands, whereas evolved No-Limit Hold'em bots combine rollout equity estimates with free-form chip raises.

\paragraph{Stationarity and unseen opponents.}
All opponents run without cross-game state and reseed their random generators from each game's deal seed so that opponent play is identical across paired runs with and without memory.
Unseen evaluation opponents differ from the training pool through held-out severity levels, inverted bet-sizing correlations, or new rule sets and evolved programs absent from both training stages.
\Cref{app:opp_examples} provides code excerpts from four unseen evaluation opponents across both games.

\subsection{Evaluation Scale and Budget}
Our evaluation spans about $51{,}800$ matches and $2.6$ million games across the two poker variants.
Every match consists of 50 sequential games against a single opponent under a context window of $L = 32{,}768$ tokens.
Our post-trained checkpoints play 128 matches per opponent both with and without memory.

Off-the-shelf models play 16 matches\textit{ per opponent} with memory in Leduc Hold'em and up to 32 in No-Limit Hold'em due to the API cost and latency of proprietary Claude and Gemini runs.
This corresponds to $15{,}200$ games with memory per model in Leduc Hold'em and up to $25{,}600$ in No-Limit Hold'em.
Single-game runs without memory never grow the context and cost far less to run.
We therefore have off-the-shelf model plays up to 128 matches \textit{per opponent }without memory to reduce variance in its baseline return.
Mean return without memory averages all completed matches, whereas experience gain uses only the matches paired with a memory run on the same card deals.

A dedicated per-game random seed fixes the card deal and stochastic opponent actions identically across paired runs.
If a model fails to produce a legal action after six retries, the harness aborts the match and discards both that match and its paired run.

\subsection{Training Details and Hyperparameters}
\label{app:training}
\label{app:sft}
\label{app:rl}

\paragraph{Supervised fine-tuning.}
We generate behavior-cloning trajectories by running Gemini~3.7 Flash \citep{comanici2025gemini} over 50-game matches with online compaction against training opponents.
Splitting recorded matches at compaction boundaries produces $3{,}746$ context-segment conversations in Leduc Hold'em and $12{,}643$ in No-Limit Hold'em.
We mask malformed reflections and decision turns that deviate sharply from both the equilibrium policy and the opponent's best response, then fine-tune one Qwen3.5-9B model per game on all remaining reasoning, action, reflection, and summary tokens.

\paragraph{Trajectory-level reinforcement learning.}
Starting from \sftm{}, we run trajectory-level GRPO using \texttt{slime} \citep{slime2025} with SGLang \citep{zheng2024sglang} for rollout generation and Megatron-LM \citep{shoeybi2019megatron} for optimization.
We run training on 48 NVIDIA H100 GPUs for Leduc Hold'em and 32 NVIDIA H100 GPUs for No-Limit Hold'em based on cluster availability.
Each training step samples 96 rollout groups across the training pool and rolls out each 20-game deal sequence $G = 12$ times in Leduc Hold'em and $G = 8$ times in No-Limit Hold'em.
This yields $1{,}152$ and $768$ matches per step, updated over eight Adam mini-batches at a constant learning rate of $2 \times 10^{-6}$ without KL or entropy regularization.
Matches that fail to output a legal action after six retries account for under $0.2\%$ of rollouts and receive a penalty reward of $-1$.
Training the final checkpoint takes about 4k H100 hours per game, and \Cref{tab:rl_hparams} summarizes all hyperparameters.

\begin{table}[h]
\centering
\footnotesize
\setlength{\tabcolsep}{10pt}
\renewcommand{\arraystretch}{1.15}
\caption{\textbf{RL hyperparameters.} Single entries apply to both games, and slash-separated entries denote Leduc Hold'em / No-Limit Hold'em.}
\label{tab:rl_hparams}
\label{tab:hparams}
\begin{tabular}{l l}
\toprule
\textbf{Parameter} & \textbf{Value} \\
\midrule
Training Opponents & 32 / 25 \\
Games per Match ($N$) & 20 \\
Rollout Groups per Step & 96 \\
Group Size ($G$) & 12 / 8 \\
Matches per Step & 1{,}152 / 768 \\
Mini-Batch Size & 144 / 96 matches \\
Optimizer & Adam ($\beta_1 = 0.9,\ \beta_2 = 0.98$, weight decay $0.1$) \\
Learning Rate & $2 \times 10^{-6}$ (constant, no warmup) \\
Gradient Norm Clip & $1.0$ \\
Clip Ratio ($\varepsilon_{\text{low}}, \varepsilon_{\text{high}}$) & $(0.20,\ 0.28)$ \\
Sampling Temperature & $1.0$ \\
Context Window ($L$) & $32{,}768$ tokens \\
Maximum Response Length & $10{,}000$ tokens \\
\bottomrule
\end{tabular}
\end{table}

\subsection{Triggering Context Compaction Between Games}
\label{app:compaction}

The harness triggers context compaction only between games so an ongoing hand is never interrupted mid-decision.
Between compactions, the buffer $H$ retains the full game transcripts, action rationales, and post-game reflections accumulated since the previous summary update.
We compact the context after a game whenever $|s| + |H|$ reaches $0.75L$, or earlier if the remaining window is smaller than the longest game observed so far in the match.
The next segment opens with the updated summary $s'$ and the six most recent one-line hand records formatted in Slumbot action-string notation \citep{jackson2017slumbot}.
Matches in which a single game overflows the window are aborted together with their paired runs in $0.5\%$ of evaluation matches.

\subsection{Details of External Feedback}
\label{app:feedback}

\paragraph{Post-game feedback tables.}
After each game, the reflection prompt presents the hand record alongside two reference tables from an equilibrium solver, illustrated in \Cref{app:prompts}.
The first table lists each decision reached by the agent, the equilibrium expected value of every legal action, and the value lost relative to equilibrium play.
The second table lists each opponent decision and the equilibrium probability of the observed move, appearing only at showdown because folded cards stay hidden.

\paragraph{Tabular CFR+ solver.}
In Leduc Hold'em, we run $2{,}000$ iterations of tabular CFR+ \citep{tammelin2014solving} in OpenSpiel \citep{lanctot2019openspiel} across all 936 information sets to reach an exploitability of $8.5 \times 10^{-5}$ chips per game.
A single tree traversal under the average policy $\pi^*$ caches the exact reach-weighted expected value of every legal action when both players follow $\pi^*$ thereafter.
After each hand, the harness replays the game history to read off these cached action values and the showdown equilibrium probabilities.

\paragraph{Monte Carlo CFR blueprint.}
Full-scale No-Limit Hold'em is too large to solve tabularly.
We therefore train an abstracted blueprint strategy for $240$ billion iterations using external-sampling Linear Monte Carlo CFR with regret-based pruning \citep{lanctot2009monte,brown2019superhuman}.
The card abstraction keeps all 169 lossless pre-flop hand classes and groups post-flop hands into $40{,}000$ flop, $20{,}000$ turn, and $8{,}000$ river buckets via $k$-means clustering on equity histograms.
Betting uses a $200$-big-blind tree of $1.01$ million decision nodes with four to six opening sizes, three to four re-raise sizes as pot fractions, and all-in.
Pseudo-harmonic action translation \citep{ganzfried2013action} maps off-tree bet sizes onto their two neighboring menu sizes across up to four weighted abstract lines.
After each hand, the harness queries the translated blueprint policy and estimates action values by Monte Carlo rollouts against the opponent's Bayesian hand range.

\subsection{Memory Structure}
\label{app:notes}

The compaction prompt instructs the model to organize past interactions into three structured entry types so that opponent observations remain separate from action plans and self-corrections.
\begin{itemize}[leftmargin=1.4em,itemsep=2pt,topsep=2pt]
\item \textbf{\texttt{READ}} entries record working hypotheses about how the opponent deviates from equilibrium, paired with a confidence level that scales with the number of observed games. Each entry records the game situation (\texttt{Scope}), supporting and conflicting hands (\texttt{Evidence}), and the observation needed to test it (\texttt{Next}).
\item \textbf{\texttt{PLAN}} entries state concrete action adjustments that exploit a recorded tendency. Each entry lists the triggering game state (\texttt{Trigger}), the supporting \texttt{READ} entry (\texttt{Based on}), and a retreat condition if opponent behavior shifts (\texttt{Guard}).
\item \textbf{\texttt{SELF}} entries record mistakes in the agent's own play surfaced during reflection, naming the relevant state (\texttt{Scope}) and the solver feedback or reasoning behind the fix (\texttt{Basis}).
\end{itemize}
At each compaction step, the model updates existing entries in place as new evidence arrives and keeps untested hypotheses marked open.

\section{Extended Experimental Results}
\label{app:results}
\label{app:full}

\subsection{Group-Level Returns and Experience Gain}
\label{app:gain_groups}

\Cref{tab:nomem} reports mean per-game returns without memory and with compacted memory across all opponent groups, complementing \Cref{tab:main}.
\Cref{tab:gain_groups} reports the corresponding paired experience gain $\Delta_{\text{exp}}$ with 95\% confidence intervals on identical card deals.
Across all opponents, \ours{} achieves the highest experience gain in Leduc Hold'em and ranks second in No-Limit Hold'em.

\noindent Comparing \Cref{tab:nomem,tab:gain_groups} clarifies the seen No-Limit Hold'em column.
Larger open-weight Qwen3.5 models show higher gains there because they start near or below zero without memory and reach at most +0.59~bb with memory.
\ours{} already earns +0.81~bb per game without memory and rises to +1.03~bb with memory.

\begin{table*}[t]
\centering
\footnotesize
\setlength{\tabcolsep}{3.1pt}
\renewcommand{\arraystretch}{1.16}
\caption{\textbf{Mean per-game returns without memory and with memory.}
Cells report mean return per game in chips for Leduc Hold'em and big blinds for No-Limit Hold'em.
Best model per column and block in bold, second-best underlined.}
\label{tab:nomem}
\resizebox{\textwidth}{!}{%
\begin{tabular}{l rrrr rrrr}
\toprule
 & \multicolumn{4}{c}{\textbf{Leduc Hold'em} {\scriptsize chips / game}} & \multicolumn{4}{c}{\textbf{No-Limit Hold'em} {\scriptsize bb / game}} \\
\cmidrule(lr){2-5}\cmidrule(l){6-9}
 & RL \& Solver & \multicolumn{2}{c}{Code-based} & Win \% & RL \& Solver & \multicolumn{2}{c}{Code-based} & Win \% \\
\cmidrule(lr){3-4}\cmidrule(lr){7-8}
\textbf{Model} & baselines {\scriptsize(6)} & Seen {\scriptsize(6)} & Unseen {\scriptsize(7)} & {\scriptsize unseen} & baselines {\scriptsize(5)} & Seen {\scriptsize(4)} & Unseen {\scriptsize(7)} & {\scriptsize unseen} \\
\midrule
\rowcolor{tabPanel}\multicolumn{9}{c}{\textbf{\textit{Without memory}}} \\
\midrule
Claude Opus 5.5 & \textbf{+0.15} & \textbf{+0.65} & \textbf{+0.23} & 62\% & \textbf{+0.30} & \textbf{+0.95} & \textbf{+1.12} & \textbf{64\%} \\
Claude Sonnet 5 & +0.13 & \underline{+0.63} & \underline{+0.22} & 60\% & +0.01 & \underline{+0.90} & +0.82 & 60\% \\
Gemini 3.1 Pro & +0.13 & +0.41 & +0.20 & \textbf{64\%} & \textminus{}0.19 & +0.13 & \textminus{}0.04 & 40\% \\
Gemini 3.7 Flash & +0.10 & +0.53 & +0.19 & \underline{63\%} & +0.13 & +0.35 & +0.40 & 53\% \\
\cmidrule(l{2pt}r{2pt}){1-9}
Qwen3.5-122B & 0.00 & +0.27 & +0.06 & 52\% & \textminus{}0.45 & +0.10 & +0.20 & 53\% \\
Qwen3.5-27B & +0.01 & +0.34 & +0.09 & 52\% & \textminus{}0.71 & \textminus{}0.13 & \textminus{}0.11 & 47\% \\
Qwen3.5-9B & \textminus{}0.18 & +0.10 & \textminus{}0.04 & 45\% & \textminus{}0.50 & 0.00 & \textminus{}0.45 & 35\% \\
\hspace{1.25em}+\,SFT & \underline{+0.14} & +0.48 & +0.18 & 61\% & \underline{+0.17} & +0.30 & +0.36 & 54\% \\
\rowcolor{tabRL}\hspace{1.25em}+\,\textbf{RL} & +0.10 & +0.46 & +0.17 & 59\% & \underline{+0.17} & +0.81 & \underline{+0.98} & \textbf{64\%} \\
\midrule
\rowcolor{tabPanel}\multicolumn{9}{c}{\textbf{\textit{With memory}}} \\
\midrule
Claude Opus 5.5 & \textbf{+0.28} & \underline{+0.91} & \textbf{+0.45} & \textbf{76\%} & \textbf{+0.64} & \textbf{+1.06} & \underline{+1.02} & \underline{65\%} \\
Claude Sonnet 5 & +0.13 & +0.63 & +0.25 & 63\% & \textminus{}0.28 & +0.89 & +0.69 & 59\% \\
Gemini 3.1 Pro & +0.12 & +0.56 & +0.24 & 70\% & \textminus{}0.16 & +0.24 & +0.20 & 50\% \\
Gemini 3.7 Flash & \underline{+0.21} & +0.80 & +0.30 & 70\% & \underline{+0.42} & +0.22 & +0.31 & 52\% \\
\cmidrule(l{2pt}r{2pt}){1-9}
Qwen3.5-122B & +0.06 & +0.52 & +0.14 & 56\% & \textminus{}0.44 & +0.59 & +0.07 & 52\% \\
Qwen3.5-27B & +0.01 & +0.50 & +0.15 & 58\% & \textminus{}0.62 & +0.38 & \textminus{}0.10 & 46\% \\
Qwen3.5-9B & \textminus{}0.19 & +0.19 & \textminus{}0.07 & 46\% & \textminus{}0.51 & \textminus{}0.63 & \textminus{}0.51 & 37\% \\
\hspace{1.25em}+\,SFT & +0.19 & +0.76 & +0.29 & 67\% & +0.27 & +0.36 & +0.42 & 57\% \\
\rowcolor{tabRL}\hspace{1.25em}+\,\textbf{RL} & +0.16 & \textbf{+0.95} & \underline{+0.40} & \underline{71\%} & \underline{+0.42} & \underline{+1.03} & \textbf{+1.28} & \textbf{66\%} \\
\bottomrule
\end{tabular}%
}
\end{table*}

\begin{table*}[t]
\centering
\footnotesize
\setlength{\tabcolsep}{3.1pt}
\renewcommand{\arraystretch}{1.18}
\caption{\textbf{Experience gain by opponent group.}
Each cell reports mean experience gain per game on paired card deals with its 95\% confidence interval in small grey type.
Best model per column in bold, second-best underlined.}
\label{tab:gain_groups}
\resizebox{\textwidth}{!}{%
\begin{tabular}{l rrrr rrrr}
\toprule
 & \multicolumn{4}{c}{\textbf{Leduc Hold'em} {\scriptsize chips / game}} & \multicolumn{4}{c}{\textbf{No-Limit Hold'em} {\scriptsize bb / game}} \\
\cmidrule(lr){2-5}\cmidrule(l){6-9}
 & \multicolumn{1}{c}{RL \& Solver} & \multicolumn{2}{c}{Code-based} & \multicolumn{1}{c}{All} & \multicolumn{1}{c}{RL \& Solver} & \multicolumn{2}{c}{Code-based} & \multicolumn{1}{c}{All} \\
\cmidrule(lr){3-4}\cmidrule(lr){7-8}
\textbf{Model} & baselines {\scriptsize(6)} & Seen {\scriptsize(6)} & Unseen {\scriptsize(7)} & opponents {\scriptsize(19)} & baselines {\scriptsize(5)} & Seen {\scriptsize(4)} & Unseen {\scriptsize(7)} & opponents {\scriptsize(16)} \\
\midrule
Claude Opus 5.5 & \textbf{+0.11}\ci{0.08} & +0.23\ci{0.10} & \textbf{+0.25}\ci{0.06} & \underline{+0.20}\ci{0.05} & \textbf{+0.35}\ci{0.57} & \textminus{}0.06\ci{0.37} & +0.19\ci{0.30} & +0.18\ci{0.24} \\
Claude Sonnet 5 & \textminus{}0.04\ci{0.06} & 0.00\ci{0.08} & +0.07\ci{0.07} & +0.01\ci{0.04} & \textminus{}0.18\ci{0.32} & \textminus{}0.09\ci{0.49} & +0.11\ci{0.36} & \textminus{}0.03\ci{0.22} \\
Gemini 3.1 Pro & \textminus{}0.03\ci{0.05} & +0.17\ci{0.05} & +0.06\ci{0.05} & +0.06\ci{0.03} & +0.06\ci{0.23} & +0.18\ci{0.25} & \underline{+0.29}\ci{0.23} & +0.19\ci{0.14} \\
Gemini 3.7 Flash & \underline{+0.08}\ci{0.06} & +0.24\ci{0.09} & +0.12\ci{0.07} & +0.15\ci{0.04} & +0.16\ci{0.26} & +0.03\ci{0.71} & \textminus{}0.03\ci{0.22} & +0.04\ci{0.22} \\
\cmidrule(l{2pt}r{2pt}){1-9}
Qwen3.5-122B & 0.00\ci{0.07} & +0.26\ci{0.10} & +0.09\ci{0.08} & +0.12\ci{0.05} & +0.13\ci{0.40} & \underline{+0.48}\ci{0.40} & +0.07\ci{0.41} & +0.19\ci{0.24} \\
Qwen3.5-27B & \textminus{}0.03\ci{0.08} & +0.15\ci{0.10} & +0.12\ci{0.08} & +0.08\ci{0.05} & +0.15\ci{0.42} & \textbf{+0.60}\ci{0.47} & +0.22\ci{0.34} & \textbf{+0.29}\ci{0.23} \\
Qwen3.5-9B & \textminus{}0.05\ci{0.10} & +0.15\ci{0.09} & 0.00\ci{0.09} & +0.03\ci{0.05} & \textminus{}0.04\ci{0.47} & \textminus{}0.68\ci{0.55} & +0.06\ci{0.47} & \textminus{}0.16\ci{0.29} \\
\hspace{1.25em}+\,SFT & +0.05\ci{0.02} & \underline{+0.28}\ci{0.03} & +0.11\ci{0.02} & +0.15\ci{0.01} & +0.11\ci{0.15} & +0.05\ci{0.20} & +0.06\ci{0.12} & +0.07\ci{0.09} \\
\rowcolor{tabRL}\hspace{1.25em}+\,\textbf{RL} & +0.06\ci{0.03} & \textbf{+0.49}\ci{0.03} & \underline{+0.23}\ci{0.03} & \textbf{+0.26}\ci{0.02} & \underline{+0.25}\ci{0.27} & +0.22\ci{0.25} & \textbf{+0.30}\ci{0.21} & \underline{+0.26}\ci{0.14} \\
\bottomrule
\end{tabular}%
}
\end{table*}

\subsection{LLMs as Opponents}
\label{app:llm_opp}

\begin{figure}[h]
\centering
\includegraphics[width=0.55\textwidth]{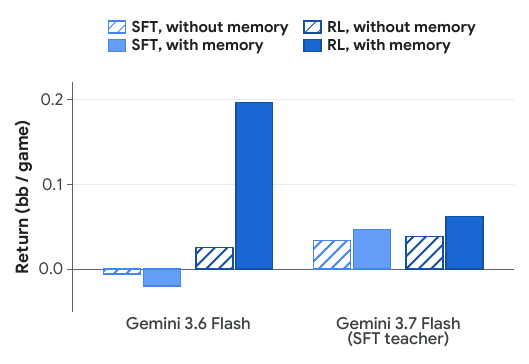}
\vspace{-0.6em}
\caption{\textbf{Returns against fixed LLM opponents in No-Limit Hold'em.}
Bars give the mean return per game on the card deals played both with and without memory.}
\label{fig:llm_opp}
\label{tab:llm_opp}
\end{figure}

We also evaluate our No-Limit Hold'em checkpoints against Gemini~3.6 Flash and Gemini~3.7 Flash as fixed opponents.
Both LLM opponents play without memory and observe only the current game so their policies remain stationary rather than introducing two-sided adaptation.
\Cref{fig:llm_opp} shows that \ours{} achieves the largest gain from memory against Gemini~3.6 Flash, whereas the remaining differences are small and have overlapping confidence intervals.

\subsection{Opponent Hand Prediction}
\label{app:solver}
\label{app:probe}

\begin{figure}[h]
\centering
\includegraphics[width=0.42\textwidth]{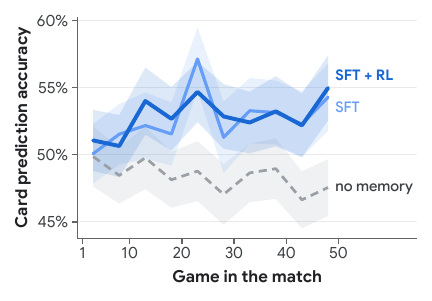}
\vspace{-0.6em}
\caption{\textbf{Predicting the opponent's hidden card in Leduc Hold'em.}}
\label{fig:analysis}
\end{figure}

We test whether higher experience gain stems from inferring hidden cards more accurately or from acting more decisively on the same beliefs.
A side query at the agent's final decision of each game asks it to predict the opponent's hidden card outside the match context.
\Cref{fig:analysis} shows that in Leduc Hold'em with memory, card-rank accuracy rises from 51\% to 55\% for \ours{}.
Accuracy for \sftm{} rises nearly identically from 50\% to 54\%, whereas play without memory stays flat near 50\%.
In No-Limit Hold'em, hand-category accuracy for \ours{} reaches $52\%$ late in the match with memory, close to $51\%$ without memory.
Supervised fine-tuning already matches RL on hidden-card prediction, showing that RL's added experience gain comes from acting on stored beliefs rather than sharpening them.

\subsection{Learning Curves Against Each Opponent}
\label{app:curves}

\begin{figure*}[t]
\centering
\includegraphics[width=\textwidth]{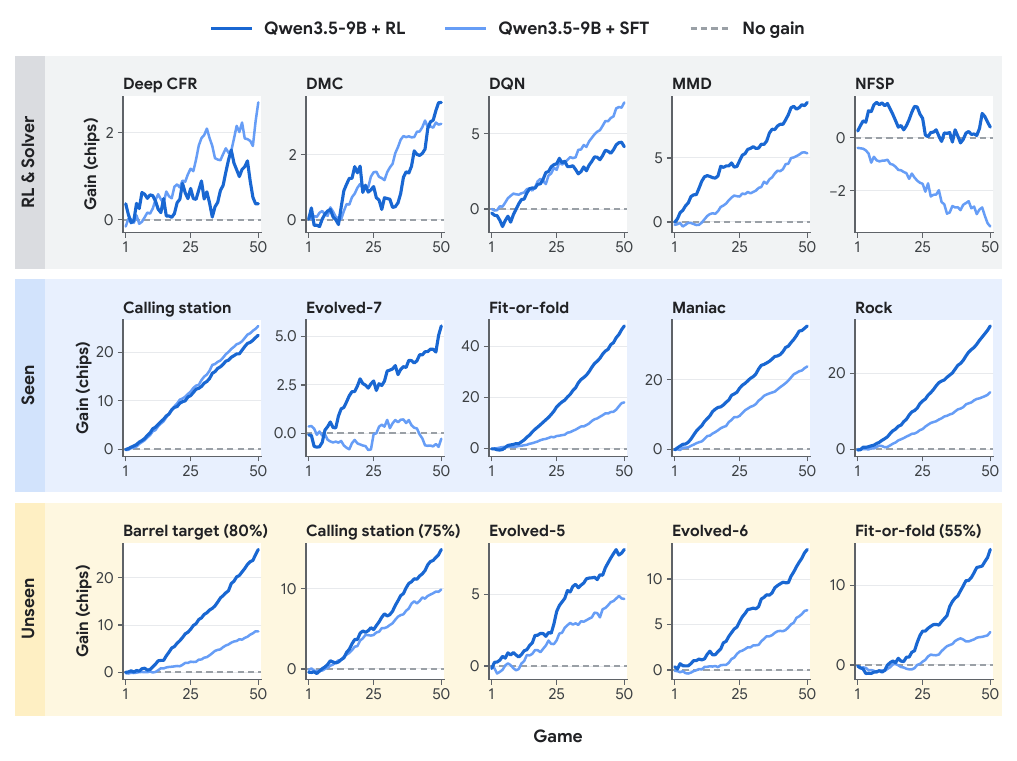}
\vspace{-1.2em}
\caption{\textbf{Cumulative experience gain per opponent in Leduc Hold'em.}
Each panel plots the cumulative return with memory minus the return without memory, in chips, on identical card deals.
Each row shows one opponent source, and rows with more than five opponents keep the five with the largest final gain of \ours{}.}
\label{fig:curve-opp-leduc}
\label{fig:curve-families}
\end{figure*}

\begin{figure*}[t]
\centering
\includegraphics[width=\textwidth]{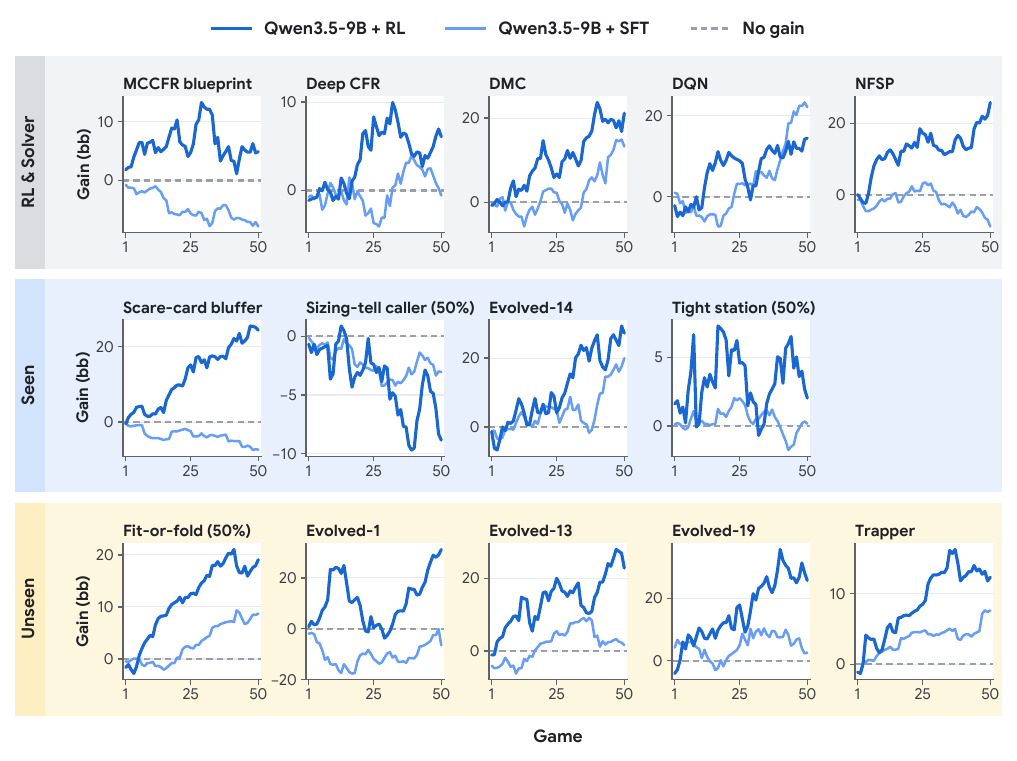}
\vspace{-1.2em}
\caption{\textbf{Cumulative experience gain per opponent in No-Limit Hold'em.}
Each panel plots the same difference as \Cref{fig:curve-opp-leduc} in big blinds.
Rows group the opponents by source as in \Cref{fig:curve-opp-leduc}, and the seen row shows all four opponents.}
\label{fig:curve-opp-hunl}
\end{figure*}

\Cref{fig:curve-opp-leduc,fig:curve-opp-hunl} plot the cumulative experience gain over a 50-game match, grouping the opponents by source and showing up to five opponents per source.
By Game~50, \ours{} has a positive cumulative gain against 18 of 19 opponents in Leduc Hold'em and 14 of 17 opponents in No-Limit Hold'em.
The sole exception in Leduc Hold'em is CFR+, whose equilibrium policy has no exploitable weakness.
\sftm{} reaches a positive cumulative gain against 16 of 19 and 11 of 17 opponents.

\clearpage
\section{Details of the Memory-Doing Gap Analysis}
\label{app:extra}
\label{app:audit}

\Cref{tab:diag} summarizes the three diagnostic metrics introduced in \Cref{sec:memory_gap} and applied to our post-trained checkpoints in \Cref{sec:rl-teaches}.
We first define the shared decision states and blind judge protocol, then detail how summaries and decisions are scored in each game.

\begin{table}[h]
\centering
\footnotesize
\setlength{\tabcolsep}{5pt}
\renewcommand{\arraystretch}{1.15}
\caption{\textbf{Overview of the diagnostic metrics.}
All metrics are computed in Leduc Hold'em against MMD, NFSP, DMC, and DQN.
Each metric scores either the final compacted summary or individual decisions in games 11 to 50, and sample sizes are averaged over the seven off-the-shelf models.}
\label{tab:diag}
\begin{tabular}{l l l l l}
\toprule
\textbf{Metric} & \textbf{Object scored} & \textbf{Reference} & \textbf{Scorer} & \textbf{Avg.\ samples per model} \\
\midrule
Identification & Final summary & Opponent profile & LLM judge & 29 summaries \\
Planning & Final summary & Opponent profile & LLM judge & 29 summaries \\
Execution & Decision & Exact best response & Computed & 1{,}290 decisions \\
\bottomrule
\end{tabular}
\end{table}

\subsection{Shared Definitions}
\label{app:audit_defs}

\paragraph{Shared decision states.}
Paired runs with and without memory play identical card deals.
A decision state $x$ is \emph{shared} when both runs follow the exact same action sequence within the current game up to $x$.
At each shared state, $a_M(x)$ and $a_{\varnothing}(x)$ denote the actions chosen with and without memory and differ only through memory and sampling.
We compute all decision-level metrics on shared states from Game~11 through Game~50 after the agent has accumulated initial experience.

\paragraph{Blind LLM judge.}
We use Claude Opus~5.5 as a blind judge for all text-based diagnostic metrics.
Items from all evaluated models are pooled in random order with model identities removed.
Decision-level prompts also omit the action chosen by the agent so the judge scores only what the summary implies.
Independently re-scoring a reshuffled $15\%$ subset reproduces at least $98\%$ of the binary judgments.

\subsection{Scoring Summaries in Leduc Hold'em}
\label{app:audit_summary}

\paragraph{Ground-truth opponent profiles.}
We construct a reference profile for each opponent from its empirical action frequencies at every information set across more than $50{,}000$ games.
From these frequencies, we extract the situations where the opponent consistently deviates from the CFR+ equilibrium.
For example, MMD never bets or calls with a Jack, whereas DQN raises every first-round bet regardless of its private card.

\paragraph{Identification and Planning.}
We sample an average of 29 matches with memory per off-the-shelf model across the four opponents and extract the compacted summary $s$ after Game~50.
Including our two post-trained checkpoints yields 252 summaries in total.
The blind judge compares each summary against the opponent's reference profile to assign two binary labels.
A summary satisfies \emph{Identification} if it records at least one exploitable tendency from the profile, and satisfies \emph{Planning} if it also states an action adjustment that exploits that tendency.

\subsection{Scoring Decisions in Leduc Hold'em}
\label{app:audit_execution}

\paragraph{Execution.}
In Leduc Hold'em, we compute the exact best response $a^{*}(x)$ against each opponent by traversing the full game tree of 936 information sets.
Let $\delta(x)$ denote how many chips $a^{*}(x)$ gains over the next-best action at $x$.
We evaluate shared states where $a^{*}(x)$ gains at least $0.1$ chips and the run without memory chooses a suboptimal action,
\begin{equation}
\mathcal{D} \;=\; \big\{\, x \;:\; \delta(x) \ge 0.1 \ \text{and}\ a_{\varnothing}(x) \neq a^{*}(x) \,\big\},
\end{equation}
which contains $1{,}290$ states per model.
Execution is the fraction of these states in which memory switches the agent to the best response,
\begin{equation}
\text{Execution} \;=\; \frac{1}{|\mathcal{D}|}\sum_{x \in \mathcal{D}} \mathbb{1}\big[\,a_M(x) = a^{*}(x)\,\big].
\end{equation}

\paragraph{Execution with no memory.}
Independent action sampling lets $a_M(x)$ match $a^{*}(x)$ even when memory has no effect.
We measure this rate for each model in Game~1, where both runs act from an empty memory $M_0 = \varnothing$.
Execution above this Game~1 baseline isolates the shift driven by memory.

\subsection{Scoring Decisions in No-Limit Hold'em}
\label{app:audit_following}

\paragraph{Sampled decision states.}
Computing an exact best response is intractable in full-scale No-Limit Hold'em, where we instead test whether each model follows its own active summary $s$.
We evaluate six off-the-shelf models because raw No-Limit Hold'em transcripts with memory for Qwen3.5-9B were not archived.
For each model and opponent, we sample five matches with memory and draw one game between Game~11 and Game~50 per match.
Taking one shared pre-flop decision and one shared post-flop decision from each sampled game yields 30 states per model and 180 in total.

\paragraph{Summary following.}
At each sampled state, the judge reads $s$ alongside the agent's hole cards, community board, pot size, and betting history.
The judge classifies the action recommended by $s$ as fold, check or call, bet under the pot, or bet at least the pot.
Bet recommendations without an explicit size count toward either bet class.
On states where $s$ makes a clear recommendation, we compare how often $a_M(x)$ and $a_{\varnothing}(x)$ fall into that class to separate summary-driven changes from default play.

\section{Example Opponent Programs}
\label{app:opp_examples}

\Cref{lst:leduc_heur,lst:leduc_evol,lst:hunl_heur,lst:hunl_evol} present core excerpts from four unseen evaluation opponents spanning both games and both construction sources.
In Leduc Hold'em, \Cref{lst:leduc_heur} shows a perturbation of $\pi^*$ that bets Queens aggressively while suppressing bluffs on other ranks, and \Cref{lst:leduc_evol} shows an evolved policy that check-raises board pairs after an opponent raise.
In No-Limit Hold'em, \Cref{lst:hunl_heur} defines a persona that bets small with strong hands and overbets when bluffing to invert a bet-sizing correlation from the training pool.
\Cref{lst:hunl_evol} shows the evolved opponent from \Cref{fig:showcase_station}, whose rollout equity never narrows the opponent's range on later rounds and therefore calls large multi-round bets.

\par\noindent\begin{minipage}{\linewidth}
\begin{lstlisting}[caption={\textbf{Leduc Hold'em hand-engineered perturbation.} Unseen opponent \texttt{q\_worshipper} applies two sequential edits over the 936-information-set CFR+ equilibrium $\pi^*$, suppressing bets outside Queens and raising aggression on Queens at severity $\alpha = 0.7$.}, label={lst:leduc_heur}]
# Unseen preset specification
"q_worshipper": _spec(
    name="q_worshipper", tier=4, base="gto", eval_only=True,
    edits=[
        _edit("no_bluff", params={"strong_r1": ["Q"], "strong_r2": ["Q"]}),
        _edit("aggro", scope={"private_rank": "Q"}, severity=0.7),
    ],
)

# Scoped tendency compiler over all 936 CFR+ information sets
policy = {k: dict(d) for k, d in gto_policy.items()}
for edit in spec["edits"]:
  fn, severity = TENDENCIES[edit["tendency"]], float(edit.get("severity", 1.0))
  for key, iset in infosets.items():
    if not _scope_matches(edit.get("scope", {}), iset):
      continue
    legal, current = legal_by_key[key], policy[key]
    target = fn(iset, legal, current, edit.get("params", {}))
    if target is not current:
      policy[key] = {
          a: (1.0 - severity) * current.get(a, 0.0) + severity * target.get(a, 0.0)
          for a in legal
      }
\end{lstlisting}
\end{minipage}\par\vspace{0.35em}

\par\noindent\begin{minipage}{\linewidth}
\begin{lstlisting}[caption={\textbf{Leduc Hold'em CSRO-evolved policy.} Round-2 logic of unseen opponent \texttt{evolved\_bot\_5}, which check-traps 55\% of board pairs after an opponent raise, checks unpaired high cards out of position, and calls single bets by pot odds.}, label={lst:leduc_evol}]
class LeducBot:
  def act(self, obs: dict[str, Any]) -> dict[str, Any]:
    my_card, community = obs["my_card"], obs["community_card"]
    call_amount, pot = int(obs["call_amount"]), int(obs["pot"])
    can_raise, position = bool(obs["can_raise"]), obs["position"]
    raises_rem = int(obs["raises_remaining"])
    pot_odds = float(call_amount) / max(1.0, float(pot + call_amount))
    if obs["round"] == "PREFLOP":
      ...  # Round-1 rank thresholds omitted for brevity

    if my_card == community:  # Board pair (strongest possible hand)
      if call_amount == 0 and position == "P0":
        lead_freq = 0.45 if self._opp_raise_count(obs) > 0 else 0.95
        if can_raise and self.rng.random() < lead_freq:
          return {"action_type": "RAISE"}
        return {"action_type": "CHECK_CALL"}
      return {"action_type": "RAISE" if can_raise else "CHECK_CALL"}

    equity = self._unpaired_equity(my_card, community)  # 0.625 (K-high) or 0.125 (J-high)
    if call_amount == 0:
      if equity < 0.30 and can_raise and self.rng.random() < 0.20:
        return {"action_type": "RAISE"}
      if equity >= 0.30 and position == "P1" and can_raise and self.rng.random() < 0.10:
        return {"action_type": "RAISE"}
      return {"action_type": "CHECK_CALL"}  # Caps P0 unpaired check range

    if raises_rem >= 1:  # Facing a single bet
      if equity > pot_odds:
        return {"action_type": "CHECK_CALL"}
      defend = 0.66 if self._opp_raise_count(obs) > 0 else 0.42
      if equity > pot_odds - 0.12 and self.rng.random() < defend:
        return {"action_type": "CHECK_CALL"}
      return {"action_type": "FOLD"}
    return {"action_type": "CHECK_CALL" if equity * 0.65 > pot_odds else "FOLD"}
\end{lstlisting}
\end{minipage}\par\vspace{0.35em}

\par\noindent\begin{minipage}{\linewidth}
\begin{lstlisting}[caption={\textbf{No-Limit Hold'em hand-engineered persona.} Unseen YAML persona \texttt{reverse\_sizing\_tell} inherits from \texttt{baseline} and inverts bet sizing by betting $0.5\times$ pot with strong hands and $1.5\times$ pot on bluffs, followed by runtime cell lookup.}, label={lst:hunl_heur}]
# Unseen persona specification
name: reverse_sizing_tell
base: baseline
postflop:
  default:
    none:                                     # Checked to or first to act
      NUTS:   {BET_S: 0.8, CHECK_CALL: 0.2}   # 0.5x pot with strongest hands
      STRONG: {BET_S: 0.8, CHECK_CALL: 0.2}
      MEDIUM: {BET_S: 0.5, CHECK_CALL: 0.5}
      WEAK:   {CHECK_CALL: 0.7, BET_L: 0.3}
      AIR:    {BET_L: 0.5, CHECK_CALL: 0.5}   # 1.5x pot overbet when bluffing
      DRAW:   {BET_L: 0.5, CHECK_CALL: 0.5}
    small: {NUTS: {BET_S: 0.7, CHECK_CALL: 0.3}}
    big:   {NUTS: {BET_S: 0.7, CHECK_CALL: 0.3}}

# Runtime cell lookup and board-override check
def decide(policy: CompiledPolicy, sit: Situation) -> dict[str, float]:
  key = (sit.tier, sit.hand.bucket) if sit.street == 0 else (
      sit.street, sit.hand.bucket, sit.hand.texture, sit.facing, sit.position
  )
  dist = policy.preflop[key] if sit.street == 0 else policy.postflop[key]
  if sit.street > 0:
    for condition, cells in policy.conditional:
      if key in cells and condition.matches(sit.hand):
        dist = cells[key]
  return dict(dist)
\end{lstlisting}
\end{minipage}\par\vspace{0.35em}

\par\noindent\begin{minipage}{\linewidth}
\begin{lstlisting}[caption={\textbf{No-Limit Hold'em CSRO-evolved policy.} Post-flop excerpt of unseen opponent \texttt{static\_bot\_13} from \Cref{fig:showcase_station}, which estimates rollout equity against a loose pre-flop filter without narrowing ranges on later rounds and calls large multi-round bets.}, label={lst:hunl_evol}]
class PokerAgent:
  def act(self, obs: dict[str, Any]) -> dict[str, Any]:
    ...  # Pre-flop logic and observation parsing omitted for brevity
    pf_raises = sum(1 for a in pf_actions if a.get("action") == "RAISE")
    bet_to_pot = call_amount / max(1.0, float(pot))
    opp_min_eq = 0.52 if pf_raises >= 2 else (0.42 if pf_raises == 1 else (
        0.48 if bet_to_pot >= 0.8 else 0.0))
    equity = self._estimate_postflop_equity(my_hand, community, 250, opp_min_eq)
    ev_call = equity * (pot + call_amount) - call_amount

    if equity >= 0.78:  # Strongest hands: size up to full pot or all-in at low SPR
      if spr < 2.2 and max_raise >= min_raise:
        return {"action_type": "RAISE", "amount": max_raise}
      sizing = 0.60 if street == "FLOP" else (0.75 if street == "TURN" else 1.0)
      target = pot + call_amount + int(sizing * pot)
      return {"action_type": "RAISE", "amount": self._clamp_raise(target, min_raise, max_raise)}

    if equity >= 0.63:  # Strong made hands: bet when checked to, mostly call bets
      if call_amount == 0 or (rand_val < 0.28 and max_raise >= min_raise):
        sizing = 0.55 if call_amount > 0 or street == "FLOP" else (
            0.60 if street == "TURN" else 0.65)
        target = pot + call_amount + int(sizing * pot)
        return {"action_type": "RAISE", "amount": self._clamp_raise(target, min_raise, max_raise)}
      return {"action_type": "CHECK_CALL"}

    if equity >= 0.48:  # Medium pairs: overestimate rollout equity on late rounds and call
      if call_amount == 0:
        ...  # Position-dependent continuation bet omitted
      if ev_call >= 0 or equity >= pot_odds:
        return {"action_type": "CHECK_CALL"}

    mdf = 1.0 - (call_amount / max(1.0, float(pot + call_amount)))
    if ev_call >= 0 or (pot_odds <= 0.25 and equity >= 0.22) or (
        equity >= pot_odds * 0.80 and rand_val < mdf * 0.40):
      return {"action_type": "CHECK_CALL"}
    return {"action_type": "FOLD"}
\end{lstlisting}
\end{minipage}\par

\section{Prompts}
\label{app:prompts}

This section provides the system prompts for both games, an example of the post-game solver feedback, and the blind judge prompts used in \Cref{app:audit}.
Each context segment opens with the system prompt and active memory summary, after which user messages supply the current game state and legal actions at each turn.

\subsection{System Prompt for Leduc Hold'em}
\label{app:prompt_leduc}

The Leduc Hold'em system prompt describes the rules, a decision framework, and the structure of the memory notes.

\begin{promptbox}[promptSys]{System prompt for Leduc Hold'em}

\promptline{You are an expert poker agent playing a long repeated match of heads-up Leduc Hold'em \textemdash{} many hands against the SAME opponent, with chips carrying over across hands. You only see your own cards during the hand.}
\promptgap
\promptline{Guiding principles:}
\promptitem{- Primary objective: maximize expected value (EV). Every decision must be justified by its overall profitability across the match.}
\promptitem{- Core strategy: GTO-informed exploitation. Use Game-Theory-Optimal play as your baseline, then deviate to exploit this opponent's specific tendencies and mistakes when you have evidence for them.}
\promptgap
\promptline{Rules: Leduc hold'em: 6-card deck (J,Q,K in two suits), 2 players, fixed-limit. Each gets one private card; one public card is revealed after round 1. Two betting rounds (bets 2 then 4, max 2 raises/round). Actions: fold/call/check/raise. Pair beats high card.}
\promptgap
\prompthead{How to decide}
\promptline{Briefly outline the situation, then reason with the concepts that apply to THIS spot:}
\promptitem{- Opponent model: what scoped range and action-policy tendencies has this opponent shown across prior hands and in your memory?}
\promptitem{- Ranges: maintain distributions rather than a single guessed hand. Start from the baseline, then reweight the opponent's holdings from position, street, and each observed action; use memory only where it supports a different weighting. Consider your own range as well.}
\promptitem{- Board: the single community card \textemdash{} does it pair either player's likely card, and whose range does it favor?}
\promptitem{- Pot odds: the price you are getting to call versus your equity.}
\promptitem{- Forward plan: if a stored hypothesis has a planned test matching this spot, decide whether this is a sufficiently low-cost opportunity. If you execute the probe, explicitly name the hypothesis and the behavior you predict observing.}
\promptitem{- Scope and uncertainty: use a read where the available evidence supports it. If you transfer a read to a meaningfully different situation, name that assumption and lower your confidence.}
\promptitem{- Read-to-action reasoning: distinguish what you observed from how you plan to adjust. Explain why the read changes the relative value of the legal actions and when the adjustment should stop applying.}
\promptitem{- Anything else you consider particularly relevant.}
\promptgap
\promptline{On a gameplay turn, before your action, briefly state:}
\promptitem{- Beliefs: the key reads/assumptions driving this decision.}
\promptitem{- Confidence: your certainty in those beliefs and in your action (high / medium / low).}
\promptline{On a hand-result turn, audit what the hand actually supports, whether it updates an existing read or hypothesis, and what should be remembered for future hands.}
\promptgap
\prompthead{Pricing aggression (fold equity)}
\promptline{A bet or raise profits from two sources: the pot taken immediately when the opponent folds, and showdown value when they continue. A raise can therefore be correct even with a hand that loses every showdown, if the opponent folds often enough. At every decision where raising is legal, price it explicitly before discarding it:}
\promptitem{- Estimate P(fold) for this opponent in this spot from your evidence \textemdash{} the OPPONENT STATS block when one is provided, otherwise your own tallies: how often have they folded when facing a bet or raise in comparable situations?}
\promptitem{- EV(raise) $\approx$ P(fold) $\times$ current pot + (1 $-$ P(fold)) $\times$ your EV when called.}
\promptitem{- Compare that against the EV of your passive options, citing the fold counts you used.}
\promptline{Beware the censored branch: you only observe how the opponent responds to a raise if you actually raise. "They never fold" is not evidence when they have rarely FACED a raise \textemdash{} count how often the branch was even tested before concluding. An unpriced or barely-tested fold branch is a standing cheap-probe candidate: plan a low-cost raise to price it (name the probe and the response you predict), rather than defaulting to passive lines forever. Weigh P(fold) against the opponent's realistic holdings in the spot, not against their strongest possible hand; showdowns and revealed folds tell you what they actually bet or raised with.}
\promptline{Keep running per-situation fold/continue tallies in your notes (e.g. "faced round-2 raise: folded 4/5") so these estimates stay grounded in observed counts, not impressions. The same arithmetic cuts both ways: against an opponent whose TESTED tallies show they genuinely rarely fold, a weak-hand raise loses its main profit source \textemdash{} prefer showdown-value lines there.}
\promptgap
\prompthead{Memory}
\promptline{Keep a compact numbered playbook for future hands. Use the note shape that matches the information being stored:}
\promptline{READ [confidence; open\textbar{}supported]: \textless{}opponent reading or working hypothesis\textgreater{} \textbar{} Scope: \textless{}position, street, and action sequence where it applies\textgreater{} \textbar{} Range: \textless{}how the observed action sequence reweights possible holdings\textgreater{} \textbar{} Evidence: \textless{}support, counterevidence, and live alternatives\textgreater{} \textbar{} Next: \textless{}what would support, refute, or resolve it\textgreater{}}
\promptline{PLAN: \textless{}intended adjustment or verification opportunity\textgreater{} \textbar{} Trigger: \textless{}spot where it applies\textgreater{} \textbar{} Based on: \textless{}supporting READ and range logic\textgreater{} \textbar{} Guard: \textless{}when to stop or not apply it\textgreater{}}
\promptline{SELF: \textless{}correction to your own play or reasoning\textgreater{} \textbar{} Scope: \textless{}where it applies\textgreater{} \textbar{} Basis: \textless{}supporting audit or reasoning\textgreater{}}
\promptline{Principles:}
\promptitem{- Range first: prefer a scoped update to the opponent's conditional range or policy over a broad personality label.}
\promptitem{- Evidence: separate observations from interpretation and adjustment. Use hand IDs and counts when available, but never invent precision.}
\promptitem{- Uncertainty: keep scope and confidence proportional to comparable evidence. Treat play as a mixed policy; an unobserved branch remains unknown, and early evidence does not justify always/never claims.}
\promptitem{- Working hypotheses: use an open READ for a decision-relevant reading that still needs verification. Preserve plausible alternatives and a falsifiable Next condition; if the test was not reached, the hypothesis remains open.}
\promptitem{- Updating: weigh recent comparable evidence alongside older evidence. If they conflict, lower confidence and keep policy change as one possible explanation; do not assume either stability or change from a small sample. Do not force a note or exploit from every hand; a strong READ or new PLAN needs diagnostic evidence or repeated comparable weak evidence. Revise rather than duplicate, and make no change when nothing durable was learned.}
\promptgap
\promptline{An OPPONENT STATS block of auto-tallied frequencies from the hands you have seen appears at the start of each context segment, stamped with the hand it is current through, and extend them yourself for hands played since. Read it critically: a 0-sample (untested) branch is missing evidence, not evidence of absence, and small samples deserve wide uncertainty. Your memory notes should carry interpretation \textemdash{} reads, range hypotheses, plans \textemdash{} rather than duplicate its raw counts.}

\end{promptbox}

\subsection{System Prompt for No-Limit Hold'em}
\label{app:prompt_hunl}

The No-Limit Hold'em system prompt follows the decision framework of Kaggle Game Arena and adds the same memory instructions as in Leduc Hold'em.

\begin{promptbox}[promptSys]{System prompt for No-Limit Hold'em}

\promptline{You are a world-class Heads-Up No-Limit Texas Hold'em (HU NLHE) poker AI. Your task is to analyze the given hand and determine the optimal play based on the following framework.}
\promptgap
\prompthead{I. Guiding Principles}
\promptgap
\promptitem{A. Primary Objective: Maximize Expected Value (EV).}
\promptline{Every decision must be justified by its overall profitability.}
\promptgap
\promptitem{B. Core Strategy: GTO-Informed Exploitation.}
\promptline{Use Game Theory Optimal (GTO) play as your baseline, but deviate to maximally exploit your opponent's specific tendencies and mistakes.}
\promptgap
\prompthead{II. Decision-Making Framework}
\promptgap
\promptitem{A. Your analysis should:}
\promptitem{- Clearly outline the current situation.}
\promptitem{- Consider key concepts in modern poker theory.}
\promptgap
\promptitem{B. Key concepts to consider:}
\promptitem{- Opponent Profile: What are their stats and tendencies (e.g., aggressive, passive, sticky, tight)?}
\promptitem{- Ranges: Who has the range and nut advantage on this board? What is my opponent's likely hand range given the action and our positions? What is my range in the current situation?}
\promptitem{- Board Texture: Is the board static/dynamic, wet/dry? How does it interact with the ranges?}
\promptitem{- Pot Dynamics: What is the Stack-to-Pot Ratio (SPR)? What are the pot odds and implied odds?}
\promptitem{- Line \& Sizing: What is my overall plan for the hand? Justify chosen bet sizes.}
\promptitem{- Game Rules: E.g., what are the minimum bets allowed in a given situation?}
\promptitem{- Anything else that you think is particularly relevant.}
\promptgap
\promptitem{B. Mandatory information to include in your response:}
\promptline{Beliefs: Explicitly state any key beliefs that factor into your decision.}
\promptline{Confidence: Rate your certainty in these beliefs and in your chosen action (e.g. high, medium, low).}
\promptline{Final Answer: \textless{}action\textgreater{} \textless{}size-if-bet-or-raise\textgreater{}}
\promptgap
\prompthead{III. Required Final Answer Format}
\promptgap
\promptline{All responses MUST start with your reasoning and conclude with the final answer.}
\promptline{The final answer MUST be on a single, final, new line.}
\promptline{The final answer line MUST be in the precise format:}
\promptline{Final Answer: \textless{}action\textgreater{} \textless{}size-if-bet-or-raise\textgreater{}}
\promptline{Where \textless{}action\textgreater{} is one of the following: fold, check, call, bet, raise}
\promptline{DO NOT include extra punctuation or text on the same line as the final answer.}
\promptgap
\promptline{Example invalid final answer lines:}
\promptline{final answer - bet 100  (missing capitalization and colon)}
\promptline{**Final Answer:** call  (extraneous punctuation)}
\promptgap
\promptline{Example valid final answer lines:}
\promptline{Final Answer: fold}
\promptline{Final Answer: check}
\promptline{Final Answer: call}
\promptline{Final Answer: bet 100}
\promptline{Final Answer: raise 100}
\promptgap
\promptline{If betting or raising, the size MUST refer to the total bet amount for the action that round.}
\promptitem{- E.g. if your opponent has bet 100, you must respond "raise 200" to indicate a raise of 100 additional chips; "raise 100" is not a valid response.}
\promptitem{- The amount MUST be in chips, not big blinds.}
\promptgap
\prompthead{Pricing aggression (fold equity)}
\promptline{A bet or raise profits from two sources: the pot taken immediately when the opponent folds, and showdown value when they continue. A bet or raise can therefore be correct even with a hand that loses every showdown, if the opponent folds often enough. At every decision where betting or raising is legal, price it explicitly before discarding it:}
\promptitem{- Estimate P(fold) for this opponent in this spot from your evidence \textemdash{} the OPPONENT STATS block when one is provided, otherwise your own tallies: how often have they folded when facing a bet or raise in comparable situations (same street, similar size relative to the pot)?}
\promptitem{- EV(bet or raise) $\approx$ P(fold) $\times$ current pot + (1 $-$ P(fold)) $\times$ your EV when called.}
\promptitem{- Compare that against the EV of your passive options, citing the fold counts you used.}
\promptline{Beware the censored branch: you only observe how the opponent responds to a bet if you actually bet. "They never fold" is not evidence when they have rarely FACED a bet \textemdash{} count how often the branch was even tested before concluding. An unpriced or barely-tested fold branch is a standing cheap-probe candidate: plan a low-cost bet to price it (name the probe and the response you predict), rather than defaulting to passive lines forever. Weigh P(fold) against the opponent's realistic holdings in the spot, not against their strongest possible hand; showdowns tell you what they actually bet, raised or called with.}
\promptline{Keep running per-situation fold/continue tallies in your notes (e.g. "faced a turn bet of about half pot: folded 4/5") so these estimates stay grounded in observed counts, not impressions. The same arithmetic cuts both ways: against an opponent whose TESTED tallies show they genuinely rarely fold, a weak-hand bet loses its main profit source \textemdash{} prefer showdown-value lines and bet larger for value there.}
\promptgap
\prompthead{How to use your memory}
\promptline{Play a solid, balanced (GTO) game by default. Before every decision, read your memory notes and the OPPONENT STATS block and ask what this opponent actually does in this spot. When they show a real pattern, deviate from the baseline to exploit it, and say in your reasoning which note or count you are acting on. Without such evidence, stay with the baseline. Keep the notes current, so the exploit follows the opponent and not a guess.}
\promptgap
\prompthead{Memory}
\promptline{Keep a compact numbered playbook for future hands. Use the note shape that matches the information being stored:}
\promptline{READ [confidence; open\textbar{}supported]: \textless{}opponent reading or working hypothesis\textgreater{} \textbar{} Scope: \textless{}position, street, and action sequence where it applies\textgreater{} \textbar{} Range: \textless{}how the observed action sequence reweights possible holdings\textgreater{} \textbar{} Evidence: \textless{}support, counterevidence, and live alternatives\textgreater{} \textbar{} Next: \textless{}what would support, refute, or resolve it\textgreater{}}
\promptline{PLAN: \textless{}intended adjustment or verification opportunity\textgreater{} \textbar{} Trigger: \textless{}spot where it applies\textgreater{} \textbar{} Based on: \textless{}supporting READ and range logic\textgreater{} \textbar{} Guard: \textless{}when to stop or not apply it\textgreater{}}
\promptline{SELF: \textless{}correction to your own play or reasoning\textgreater{} \textbar{} Scope: \textless{}where it applies\textgreater{} \textbar{} Basis: \textless{}supporting audit or reasoning\textgreater{}}
\promptline{Principles:}
\promptitem{- Range first: prefer a scoped update to the opponent's conditional range or policy over a broad personality label.}
\promptitem{- Evidence: separate observations from interpretation and adjustment. Use hand IDs and counts when available, but never invent precision.}
\promptitem{- Uncertainty: keep scope and confidence proportional to comparable evidence. Treat play as a mixed policy; an unobserved branch remains unknown, and early evidence does not justify always/never claims.}
\promptitem{- Working hypotheses: use an open READ for a decision-relevant reading that still needs verification. Preserve plausible alternatives and a falsifiable Next condition; if the test was not reached, the hypothesis remains open.}
\promptitem{- Updating: weigh recent comparable evidence alongside older evidence. If they conflict, lower confidence and keep policy change as one possible explanation; do not assume either stability or change from a small sample. Do not force a note or exploit from every hand; a strong READ or new PLAN needs diagnostic evidence or repeated comparable weak evidence. Revise rather than duplicate, and make no change when nothing durable was learned.}
\promptgap
\promptline{An OPPONENT STATS block of auto-tallied frequencies from the hands you have seen appears at the start of each context segment, stamped with the hand it is current through, and extend them yourself for hands played since. Read it critically: a 0-sample (untested) branch is missing evidence, not evidence of absence, and small samples deserve wide uncertainty. Your memory notes should carry interpretation \textemdash{} reads, range hypotheses, plans \textemdash{} rather than duplicate its raw counts.}

\end{promptbox}

\subsection{Post-Game Solver Feedback}
\label{app:prompt_feedback}

The box below shows the solver feedback returned after Game~1 of a Leduc Hold'em match between Qwen3.5-9B and DMC without the trailing reflection instructions.

\begin{promptbox}[promptFb]{Post-game solver feedback in Leduc Hold'em}
\begin{verbatim}
*** HAND 1 of 50 RESULT AND REFLECTION ***
Hand record:
Hand 1: You (acts first) Ks vs Opp Jh on Qh - You win 7 chips at showdown  [b2c/b4c]

Your card: King of Spades (Ks)
Opponent's card: Jack of Hearts (Jh)
Community card: Queen of Hearts (Qh)
Outcome: You win 7 chips at showdown.

GTO action-value table (chips; * marks the sampled action):
| # | Decision       | Fold | Check/Call | Bet/Raise | Declared policy | Policy EV | Cost vs GTO |
|---|----------------|------|------------|-----------|-----------------|-----------|-------------|
| 1 | Pre-flop, open | -    | +1.0       | +1.0 *    | bet 1.00        | +1.04     | 0.00        |
| 2 | Flop, open     | -    | -2.0       | -2.9 *    | bet 1.00        | -2.90     | 0.86        |

Opponent GTO action table:
| # | Opponent decision | Observed action | GTO probability | GTO mix                          |
|---|-------------------|-----------------|-----------------|----------------------------------|
| 1 | Pre-flop, vs bet  | call            | 12.6%           | fold 81.9%/call 12.6%/raise 5.5% |
| 2 | Flop, vs bet      | call            | 0.0%            | fold 100.0%/call 0.0%/raise 0.0% |
\end{verbatim}
\end{promptbox}

\subsection{Judge Prompt for Identification and Planning}
\label{app:prompt_judge_summary}

The judge compares each final match summary in Leduc Hold'em against the opponent's reference behavioral profile to score Identification and Planning.

\begin{promptbox}[promptJudge]{Judge prompt for Identification and Planning}

\promptline{You are evaluating the memory summary that a poker agent wrote during a repeated match of heads-up Leduc Hold'em against one fixed opponent.}
\promptgap
\prompthead{Opponent profile (ground truth)}
\promptline{\ph{opponent\_profile}}
\promptgap
\prompthead{Agent's summary}
\promptline{\ph{summary}}
\promptgap
\prompthead{Instructions}
\promptline{A tendency claim is a general statement about how the opponent acts in a stated situation, or about which cards an opponent action implies. A single hand reported without generalization, a style word with no situation (e.g. "aggressive"), and statements about the agent's own play are not tendency claims. Check every tendency claim against the opponent profile.}
\promptgap
\promptitem{1. Identification: does the summary state at least one exploitable tendency that the profile confirms?}
\promptitem{2. Planning: does the summary also give a concrete instruction for the agent's own play, naming an action in a specific situation, that is the profitable response to a confirmed tendency? Generic advice such as "value bet strong hands", or a plan that points in the wrong direction, does not count.}
\promptgap
\promptline{Return a JSON object:}
\promptline{\{"identification": true \textbar{} false, "planning": true \textbar{} false, "evidence": "\textless{}the confirmed claim and the plan you relied on\textgreater{}"\}}

\end{promptbox}

\subsection{Judge Prompt for Summary Following}
\label{app:prompt_judge_following}

In No-Limit Hold'em, the judge reads the summary and the decision state without the agent's action and records which action the summary implies.

\begin{promptbox}[promptJudge]{Judge prompt for summary following in No-Limit Hold'em}

\promptline{You are evaluating whether a poker agent's memory summary implies a specific action at one decision in heads-up No-Limit Hold'em. You do not see the action the agent took.}
\promptgap
\prompthead{Agent's summary}
\promptline{\ph{summary}}
\promptgap
\prompthead{Decision state}
\promptline{Opponent: \ph{opponent}    Hole cards: \ph{hole\_cards}    Board: \ph{board}    Pot: \ph{pot}    To call: \ph{to\_call}}
\promptline{Betting history: \ph{betting\_history}}
\promptgap
\prompthead{Instructions}
\promptitem{1. Applies: does the summary contain an observation or plan about this opponent whose scope covers this decision, meaning the same betting round, the same role, and the agent's hand class when the note names one? Notes about the agent's own mistakes, general poker advice, and style words with no situation do not apply.}
\promptitem{2. Action class: if the summary applies, which action does it imply here? Choose one of fold, check or call, bet smaller than the pot, bet of at least the pot, bet with no size stated, or none when the summary does not settle the action or two entries conflict.}
\promptgap
\promptline{Return a JSON object:}
\promptline{\{"applies": true \textbar{} false, "action\_class": "fold" \textbar{} "check\_call" \textbar{} "bet\_small" \textbar{} "bet\_large" \textbar{} "bet" \textbar{} "none"\}}

\end{promptbox}

\end{document}